\documentclass[pra,amsmath,amssymb,showcontents,floatfix,10pt,twocolumn,longbibliography,superscriptaddress,nofootinbib]{revtex4}
\usepackage{bm}
\usepackage{txfonts}
\usepackage{graphicx}
\usepackage{subfigure}
\usepackage{comment}
\usepackage{ifpdf}
\ifpdf
\usepackage{epstopdf}
\fi
\usepackage{color}

\usepackage{bm}
\usepackage{enumerate}
\usepackage{dsfont}
\usepackage{siunitx}

\usepackage[unicode]{hyperref}
\hypersetup{
    pdftitle={Intracavity triplet down conversion},
    pdfauthor={Mathew Denys and Ashton Bradley},
    colorlinks=true,
    linkcolor=blue,
    citecolor=blue,
    filecolor=black,
    urlcolor=blue
}
\def\urlprefix{}
   \def\url#1{}

\newcommand{\fref}[1]{Fig.~\ref{#1}}

\newcommand{\eref}[1]{(\ref{#1})}

\newcommand{\dfdx}[2]{\frac{\partial#1}{\partial#2}}

\newcommand{\av}[1]{\langle#1\rangle}

\newcommand{\kap}{\kappa}
\newcommand{\gam}{\gamma}

\begin{document}

\title{Steady states, squeezing, and entanglement in intracavity triplet down conversion}

\author{M. D. E. Denys}
\affiliation{Centre for Quantum Science, Department of Physics, University of Otago, Dunedin, New Zealand}
\affiliation{The Dodd-Walls Centre for Photonic and Quantum Technologies, New Zealand}
\author{M. K. Olsen}
\affiliation{School of Mathematics and Physics, University of Queensland, Brisbane, Queensland 4072, Australia}
\date{\today}
\author{L. S. Trainor}
\affiliation{Centre for Quantum Science, Department of Physics, University of Otago, Dunedin, New Zealand}
\affiliation{The Dodd-Walls Centre for Photonic and Quantum Technologies, New Zealand}
\author{H. G. L. Schwefel}
\affiliation{Centre for Quantum Science, Department of Physics, University of Otago, Dunedin, New Zealand}
\affiliation{The Dodd-Walls Centre for Photonic and Quantum Technologies, New Zealand}
\author{A. S. Bradley}
\affiliation{Centre for Quantum Science, Department of Physics, University of Otago, Dunedin, New Zealand}
\affiliation{The Dodd-Walls Centre for Photonic and Quantum Technologies, New Zealand}

\begin{abstract}
Triplet down conversion, the process of converting one high-energy photon into three low-energy photons, may soon be experimentally feasible due to advances in optical resonator technology.
We use quantum phase-space techniques to analyse the process of degenerate intracavity triplet down conversion by solving stochastic differential equations within the truncated positive-\textit{P} representation. The time evolution of both intracavity mode populations are simulated, and the resulting steady-states are examined as a function of the pump intensity. Quantum effects are most pronounced in the region immediately above the semi-classical pumping threshold, where our numerical results differ significantly from semi-classical predictions. Regimes of measurable squeezing and bipartite entanglement are identified from steady-state spectra of the cavity output fields. We validate the truncated positive-\textit{P} description against Monte Carlo wave function simulations, finding good agreement for low mode populations.

\end{abstract}

\maketitle

\section{Introduction}
Intracavity pair down conversion, whereby pump photons are parametrically converted into pairs of photons inside a $\chi^{(2)}$ nonlinear medium, is a well studied process due to its many applications, particularly for allowing widely-tunable light sources---even as wide as two octaves \cite{Olsen_2017,kolker_widely_2018,Olsen:2018gr,Li_2018}---and squeezed light generation \cite{furst_quantum_2011,Peano_2015,andersen_30_2016}.
The degenerate case, $2\omega\rightarrow\omega+\omega$, where the generated photons are of the same polarization, is notable for being the leader in squeezed light generation, with a record of more than $15~\mathrm{dB}$ squeezing \cite{vahlbruch_detection_2016}.
Above the optical parametric oscillation threshold, when spontaneous pair production is frequent enough to seed a large-scale transfer of energy between the modes, the same system still exhibits squeezing \cite{furst_quantum_2011}, as well as correlation in the output beams \cite{takei_high-fidelity_2005,jing_experimental_2003}. Degenerate triplet down conversion, $3\omega\rightarrow\omega+\omega+\omega$, is the third-order ($\chi^{(3)}$) equivalent of pair down conversion, and the inverse of third-harmonic generation~\cite{Cavanna:16}. This process has not yet been observed experimentally, neither above nor below threshold.
Emerging resonator technologies combining low losses with large optical overlaps and nonlinearities are bringing experimental realization closer to fruition, motivating deeper theoretical investigation.

In this work we use a truncated form of Drummond and Gardiner's positive-\textit{P} representation~\cite{Drummond_1980} to degenerate triplet down conversion in quantum phase space. The resulting equations are solved numerically to determine the time evolution of the intracavity fields of a damped, driven optical cavity, shown schematically in Fig.~\ref{fig:cavity}. Of particular interest are the resulting steady-state fields, which are examined over a range of pump intensities, including a comparison to analytical semi-classical predictions \cite{Bajer,Felbinger}. We observe good agreement except in the region immediately above the semi-classical pumping threshold where critical fluctuations are signficant. We examine this region for a range of initial conditions, and weak injected signals in the low-energy mode. We use a linearized fluctuation analysis to numerically identify regimes of squeezing and bipartite entanglement in the output fields. We perform a comparison of the time evolution given by our phase space methods with that of a more exact Monte Carlo wave function simulation in a number state basis. The results demonstrate the validity of the truncated positive-\textit{P} method for this system.

The structure of this paper is as follows: in Sec.~\ref{sys} we develop the equations of motion for truncated positive-$P$ and semiclassical approaches, and present the known semi-classical steady state solutions. In Sec.~\ref{timeevo} we estimate experimental nonlinearities to justify our choice of simulations parameters, and solve for steady states of the system comparing positive-$P$ and semiclassical approaches. In Sec.~\ref{sqz} we present numerical results for measures of squeezing and entanglement in the system, and in Sec.~\ref{disc} we discuss our results and conclude.

\section{System and Equations of Motion}\label{sys}

\begin{figure}
	\centering
	\includegraphics[width=0.9\linewidth]{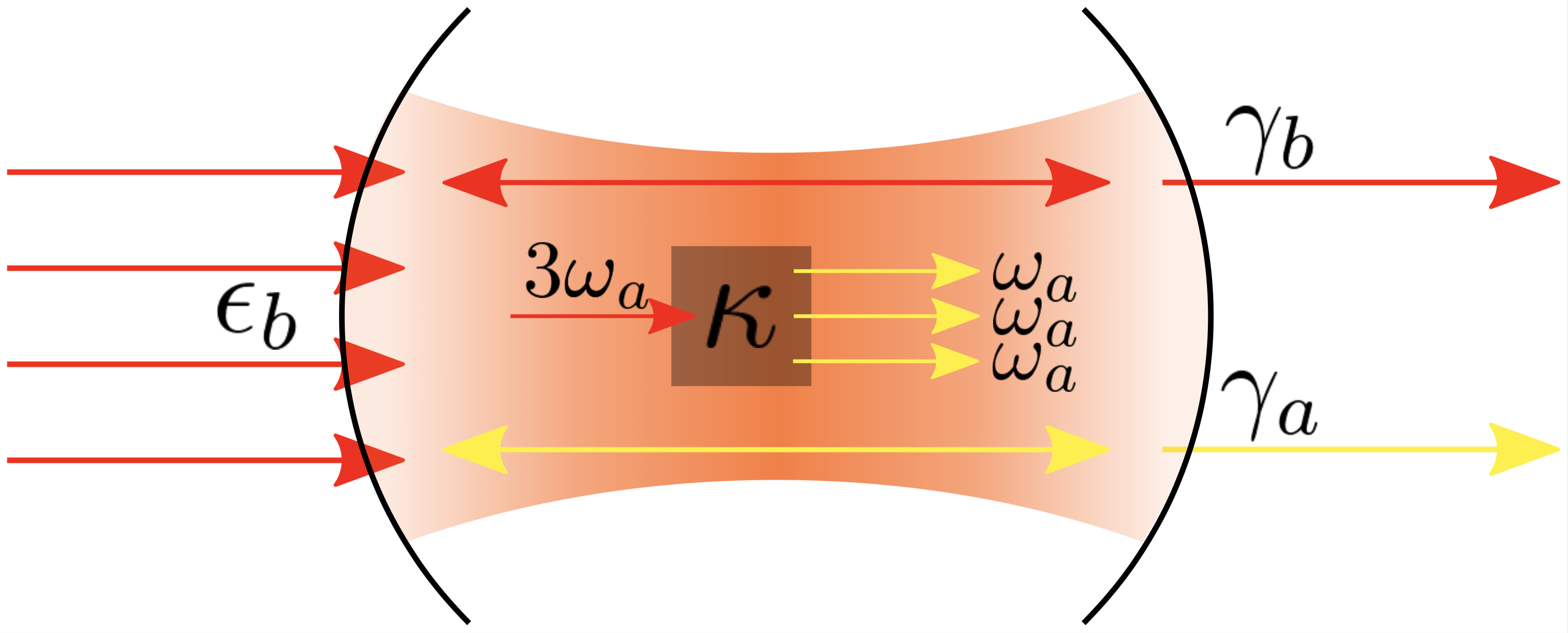}
	\caption{Our model involves an optical cavity with two occupied resonant modes, at frequencies $\omega_a$ and $\omega_b=3\omega_a$. The effective non-linearity $\kap$ allows for conversion of photons between the two modes. The cavity is coupled to the environment via (i) classical pumping into the high energy mode, and (ii) loss of photons at the rates $\gam_a$ and $\gam_b$ for the respective modes.}
	\label{fig:cavity}
\end{figure}

The system we consider is that of degenerate intracavity triplet down conversion, shown schematically in \fref{fig:cavity}. The interaction is described by the Hamiltonian
\begin{align}
\hat{H}_\text{int} = i\hbar\frac{\kap}{3} \left[\hat{a}^{\dagger 3}\hat{b}-\hat{a}^3\hat{b}^\dagger\right], \label{eq:intHamiltonian}
\end{align}
where $\hat{b}^\dagger$, $\hat{b}$ [$\hat{a}^\dagger$, $\hat{a}$] are the bosonic creation and annihilation operators for the high [low] energy photons, and $\kap\propto\chi^{(3)}$ is the effective non-linearity of the cavity medium, which can be taken to be real and positive with no loss of generality. The high-energy mode is driven by a classical laser field described by the additional Hamiltonian
\begin{align}
\hat{H}_\mathrm{pump} = i\hbar\left[\epsilon_b\hat{b}^\dagger - \epsilon_b^*\hat{b}\right],\label{eq:pump}
\end{align}
where $\epsilon_b$ is the pump amplitude of the high-energy field. Loss of both high and low energy photons from the cavity is represented by the Liouvillian superoperator acting on the system density matrix
\begin{align}
\mathcal{L}\,\rho = \gam_a [2\hat{a} \rho \hat{a}^\dagger - \hat{a}^\dagger \hat{a} \rho - \rho \hat{a}^\dagger \hat{a}]
+ \gam_b [2\hat{b} \rho \hat{b}^\dagger - \hat{b}^\dagger \hat{b} \rho - \rho \hat{b}^\dagger \hat{b}],
\end{align}
where $\gam_a$ [$\gam_b$] is the cavity loss rate of the low [high] energy mode. The overall time evolution of the density matrix is given by the Linblad master equation \cite{lindblad1976,DaleyReview}
\begin{align} \label{eq:master}
\dot{\rho} = -\frac{i}{\hbar}[\hat{H}_\mathrm{sys},\rho] + \frac{1}{2}\mathcal{L}\,\rho,
\end{align}
where $\hat{H}_\mathrm{sys} = \hat{H}_\mathrm{int} + \hat{H}_\mathrm{pump}$. Using the positive-\textit{P} operator correspondences \cite{QuantumOptics} we map \eqref{eq:master} onto a partial differential equation for the positive-\textit{P} distribution
\begin{align} \label{eq:FPE}
\begin{split}
\frac{\partial P}{\partial t} = \Bigg\{
&-\Bigg\lbrack
\frac{\partial}{\partial\alpha}(\kap\alpha^{\!+2}\beta)
+\frac{\partial}{\partial\alpha^{\!+}}(\kap\alpha^2\beta^+)
+\frac{\partial}{\partial\beta}\left(-\frac{\kap}{3}\alpha^3\right)\\
&+\frac{\partial}{\partial\beta^+}\left(-\frac{\kap}{3}\alpha^{\!+3}\right)
-\frac{\partial}{\partial\alpha}(\gam_a\alpha)
- \frac{\partial}{\partial\alpha^{\!+}}(\gam_a\alpha^{\!+})\\
&+ \frac{\partial}{\partial\beta} (\epsilon_b)
+ \frac{\partial}{\partial\beta^+} (\epsilon_b^*)
\Bigg\rbrack\\
&+\frac{1}{2}\Bigg\lbrack
\frac{\partial^2}{\partial\alpha^2}(2\kap\alpha^{\!+}\beta)
+\frac{\partial^2}{\partial\alpha^{\!+2}}(2\kap\alpha\beta^+)
\Bigg\rbrack\\
&-\frac{1}{6}\Bigg\lbrack
\frac{\partial^3}{\partial\alpha^3}(2\kap\beta^+)
+\frac{\partial^3}{\partial\alpha^{\!+3}}(2\kap\beta)
\Bigg\rbrack
\Bigg\}\\
\times& \ P(\alpha,\alpha^{\!+}\!,\beta,\beta^+),
\end{split}
\end{align}
where the positive-\textit{P} variables $\alpha$ and $\alpha^{\!+}$ are independent and complex-valued. They correspond to the operators $\hat{a}$ and $\hat{a}^\dagger$ in the sense that expectation values of normally ordered operator moments can be calculated via $\av{\hat{a}^{\dagger m}\hat{a}^n} = \int d^2\alpha \ d^2\alpha^{\!+} \ \alpha^{\!+m}\alpha^n P(\alpha,\alpha^{\!+}\!,\beta,\beta^+)$, and analogously for the $b$ mode. Due to the cubic terms in \eqref{eq:intHamiltonian}, third-order derivatives exist in \eqref{eq:FPE}, meaning it cannot be interpreted as a true Fokker--Plank equation (FPE). However, a system size expansion of \eqref{eq:FPE} shows that the third-order terms scale with $n^{-1/2}$, where $n$ is the number of photons in the cavity. We can hence reasonably expect these terms to become negligible for large mode populations. This approximation is further examined in Appendix \ref{sec:MCWF}. Truncating the third-order terms from \eqref{eq:FPE} results in a true FPE. Since the diffusion matrix is positive definite, the FPE can be mapped to a set of coupled stochastic differential equations (SDEs) using the rules of It\^{o} calculus:
\begin{align} \label{eq:ito}
\begin{split}
d\alpha       & = \left( -\gam_a\alpha   + \kap\alpha^{\!+2}\beta \right) dt + \sqrt{2\kap \alpha^{\!+}\beta}\;dW_1(t), \\
d\alpha^{\!+} & = \left( -\gam_a\alpha^{\!+} + \kap\alpha^2\beta^+ \right) dt  + \sqrt{2\kap \alpha\beta^+}\;dW_2(t),     \\
d\beta        & = \left( \epsilon_b   - \gam_b\beta   - \frac{\kap}{3}\alpha^3 \right) dt,                               \\
d\beta^+      & = \left( \epsilon_b^* - \gam_b\beta^+ - \frac{\kap}{3}\alpha^{\!+3} \right)dt,
\end{split}
\end{align}
where the Wiener increments $dW_i$ are Gaussian distributed with $\overline{dW_i(t)}=0$ and $\overline{dW_i(t) dW_{\!j}(t)}=dt$. Stochastically averaging over trajectories of \eqref{eq:ito} yields moments approximately consistent with \eqref{eq:FPE}, i.e. $\overline{\alpha^{\!+m}\alpha^n} \rightarrow \av{\hat{a}^{\dagger m}\hat{a}^n}$.
These It\^{o} SDEs are related to a set of equivalent Langevin equations, which can be obtained by dividing both sides of \eqref{eq:ito} by $dt$. The benefit of the It\^{o} SDEs is that the Wiener increments are mathematically rigourously defined, allowing for well defined numerical integration. Each of the terms present has an easily interpretable physical meaning. The $\epsilon$ terms correspond to a constant pump, which continuously increases the population of the $b$ mode. The $\gamma$ terms generate damping by removing photons from each mode at a rate proportional to the mode population. The last non-stochastic term in each equation describes the change in the fields due to the $\chi^{(3)}$ interaction, proportional to the effective interaction strength $\kappa$. The interaction leads to a decrease in the high energy mode population, and an increase in the low energy mode. We lack a factor of $\frac{1}{3}$ in the $\alpha$ and $\alpha^{\!+}$ equations because down conversion of one $b$ photon leads to three $a$ photons. The appearance of e.g. the product  $\kappa\alpha \beta^+$ in the stochastic term indicates that the stochisticity originates from the quantum nonlinear interaction.

It should be noted that, using the techniques developed in \cite{StochasticDifferenceEquations}, it is possible to develop stochastic difference equations corresponding to \eqref{eq:FPE}, which, although still approximate, do not require truncation of third-order terms. However, these equations are unsatisfactory for practical use, as convergence cannot be proven and they suffer from severe numerical instabilities \cite{ThreePhotonProcesses}.

A further, semi-classical, approximation is obtained by neglecting the noise terms in \eqref{eq:ito}. The convenient choice of initial condition $\alpha^{\!+}(0)=\alpha_0^*$ [$\beta^+(0)=\beta_0^*$], where $\alpha_0=\alpha(0)$ [$\beta_0=\beta(0)$], yields the coupled ordinary differential equations (ODEs)
\begin{align} \label{eq:ode}
\begin{split}
\dfdx{\alpha}{t}  &= \kap\alpha^{*2}\beta - \gam_a\alpha,\\
\dfdx{\beta}{t}   &= \epsilon_b - \gam_b\beta  - \frac{\kap}{3}\alpha^3,
\end{split}
\end{align}
with $\alpha^{\!+}$ recovered via $\alpha^* \to \alpha^{\!+}$, and analagously for $\beta^+$. The steady-states of \eqref{eq:ode} have been well studied analytically as a function of $\epsilon_b$ in the physically relevant $\gam_a\ll\gam_b$ regime \cite{Bajer, Felbinger}. The trivial solution $\alpha=0$, $\beta=\epsilon_b/\gam_b$ is a steady-state at all pump intensities. Steady state solutions with non-zero down converted fields are only realized above the pumping threshold $\epsilon_{b,\mathrm{th}} = 4(\gam_a\gam_b)^{3/4} / (3\!\sqrt{\kap})$. The magnitude of these non-trivial solutions is determined by solving
\begin{align} \label{eq:ode_ss_a}
|\alpha_s|^4 - \frac{3|\epsilon_b|}{\kap}|\alpha_s| + \frac{3\gam_a\gam_b}{\kap^2} = 0,
\end{align}
and each solution is triply degenerate in phase, which is independently determined by solving $e^{i3\theta} = \epsilon_b/|\epsilon_b|$. For a given set of parameters above the pumping threshold there are two realisable solutions to \eqref{eq:ode_ss_a}, which take the form of a pair of diverging branches when plotted against $\epsilon_b$. The lower branch is unstable, while the upper branch is stable for $\gam_a<\gam_b$ \cite{Felbinger}. For each steady-state solution for $\alpha$, the corresponding solution for $\beta$ is given by
\begin{align} \label{eq:ode_ss_b}
\beta_s = \frac{\epsilon_b - \kap\alpha_s^3/3}{\gam_b},
\end{align}
determined by the balance between pumping, damping, and down/up conversion.

\section{Time Evolution and Steady States}\label{timeevo}
\subsection{Experimental considerations}\label{expc}
We have to make a choice for the regime for our simulations, specifically the ration of damping to nonlinear interaction; this ratio sets the size of signficant quantum effects caused by the triplet down conversion. We motivate our choice of interaction physically by providing an estimate for $\kappa$ in a monolithic resonant cavity. Assume the fields are linearly polarized within a resonant cavity of length $L$ entirely made of a nonlinear material with effective nonlinearity $\chi^{(3)}$ and permittivity $\varepsilon_a$ [$\varepsilon_b$] for the low- [high-]energy mode. We can then perform a macroscopic quantisation of the displacement fields \cite{drummond_quantum_2014}, and substitute the quantized fields into the interaction Hamiltonian. We find for $\kappa$:
\begin{align} \label{eq:kappa}
	\kappa = \frac{3\hbar\varepsilon_0\chi^{(3)}\sqrt{\omega_a^3\omega_b}}{4 \sqrt{\varepsilon_a^3 \varepsilon_b}}\frac{\delta[m_b-3m_a]}{L}\sigma,
\end{align}
where $\varepsilon_0$ is the vacuum permittivity, $m_a$ [$m_b$] is the number of field oscillations of the low- [high-]energy mode in one round trip, and $\sigma$ is the transverse modal overlap given by
\begin{align}\label{eq:overlap}
\sigma = \frac{\int u_a^3(x,y)u_b^*(x,y)dxdy}{\left(\int|u_a(x,y)|^2 dxdy\right)^{\frac{3}{2}} \left(\int|u_b(x,y)|^2 dxdy\right)^{\frac{1}{2}}}.
\end{align}
Here $u_a$ [$u_b$] is the transverse mode profile of the displacement field of the low- [high-] energy mode, which is assumed to be constant along the length of the cavity. The Kronecker-delta term $\delta[m_b-3m_a]/L$ originates under this assumption from the overlap of the mode profiles in the direction of propagation, and asserts that the modes are phasematched.

From the most efficient recent demonstration of doubly-resonant third-harmonic generation we are aware of \cite{surya_efficient_2018}, we can extract the following parameters: $\omega_a = 2\pi\times\SI{194}{\tera\hertz}$; $L = 2\pi\times\SI{20}{\micro\meter}$; $\sigma=\SI{0.43}{\per\square\micro\meter}$ \cite{surya_private_2019}; $\chi^{(3)}\approx\SI{1.5e-20}{\square\meter\per\square\volt}$, taking the average of values for silicon nitride \cite{ning_third-harmonic_2013} and aluminium nitride \cite{mateen_micromechanical_2018}; and $\varepsilon_{a,b}\approx 4\varepsilon_0$~\cite{kischkat_mid-infrared_2012}.
From these parameters, we can estimate that $\kappa$ is of the order of magnitude of \SI{100}{\per\second}, which is seven orders of magnitude smaller than their measured loss rate of $\gamma_a=\omega_a/(2Q_a)=\SI{1.5e9}{\per\second}$, where $Q_a$ is the $Q$-factor of the low-energy mode. However, numerically we also have a limitation posed by finite resources. Thus, in all of the presented results we set $\kappa = 0.001\gamma_a$, so that the interaction remains much weaker than dissipation, but its effects can be more easily studied in simulations. A further advantage is that the artifacts of truncation that we consider further in Appendix  \ref{sec:MCWF} are exaggerated by this choice (and our choice of lower photon number), offering a pessimistic assessment of the vality of truncation.

\subsection{Simulations}
We use the adaptive LambaEulerHeun \cite{StocDiffEq} algorithm implemeted in the Julia \cite{Bezanson:2017gd} package  DifferentialEquations.jl \cite{DiffEq} to simulate the truncated positive-\textit{P} SDE \eqref{eq:ito}. Expectation values of the mode populations $\av{\hat{n}_a} = \av{\hat{a}^{\dagger}\hat{a}}$ and $\av{\hat{n}_b} = \av{\hat{b}^{\dagger}\hat{b}}$ were calculated by averaging over $10^6$ trajectories, which was found to be sufficient for convergence of the second-order moments used to characterize the system throughout this work. For comparison with a semi-classical treatment, mode populations were obtained from the mean-field equations \eqref{eq:ode} using an ODE solver.

We begin by examining the time evolution of the mode populations for typical experimental parameters. We set the initial state to be the vacuum, since for optical photons the thermal occupation of the modes will be negligible. Due to the nature of the positive-\textit{P} function as a pseudo-probability distribution, there are an infinite number of corresponding choices of distributions for $\bm{\alpha}\equiv(\alpha,\alpha^{\!+},\beta,\beta^+)$. The simplest positive-\textit{P} distribution for the vacuum is a coherent state with zero amplitude, and the distribution is a delta function at $\bm{\alpha}=\mathbf{0}$. Inspection of \eqref{eq:ito} shows that down conversion will not proceed in this case because the noise term in $\alpha$ [$\alpha^{\!+}$] is proportional to the product $\alpha^{\!+}\beta$ [$\alpha\beta^+$], which will always be zero.
We might guess that a different representation of the vacuum that involves distribution including $\bm{\alpha}\neq\mathbf{0}$ \cite{sampling} could allow for a mechanism that seeds spontaneous triplet production. However, find numerically that this is not the case, consistent with the fact that the physical result should be independent of the particular positive-\textit{P} representation. The true source of the inhibition of spontaneous processes is that the truncation of third order derivatives in \eqref{eq:FPE} is not valid for small populations, due to the invalidity of the scaling argument given above. Thus, we are unable to describe true spontaneous processes within a truncated positive-\textit{P} approach. Inclusion of the third-order terms would allow for spontaneous triplet production \cite{ThreePhotonProcesses}, but at prohibitive technical cost. In contrast, the positive-\textit{P} SDE for spontaneous pair production requires no truncation, and has noise terms proportional to only $\beta$ and $\beta^+$, which seed the process \cite{SHG}.

\begin{figure}
	\centering
	\includegraphics[width=\linewidth]{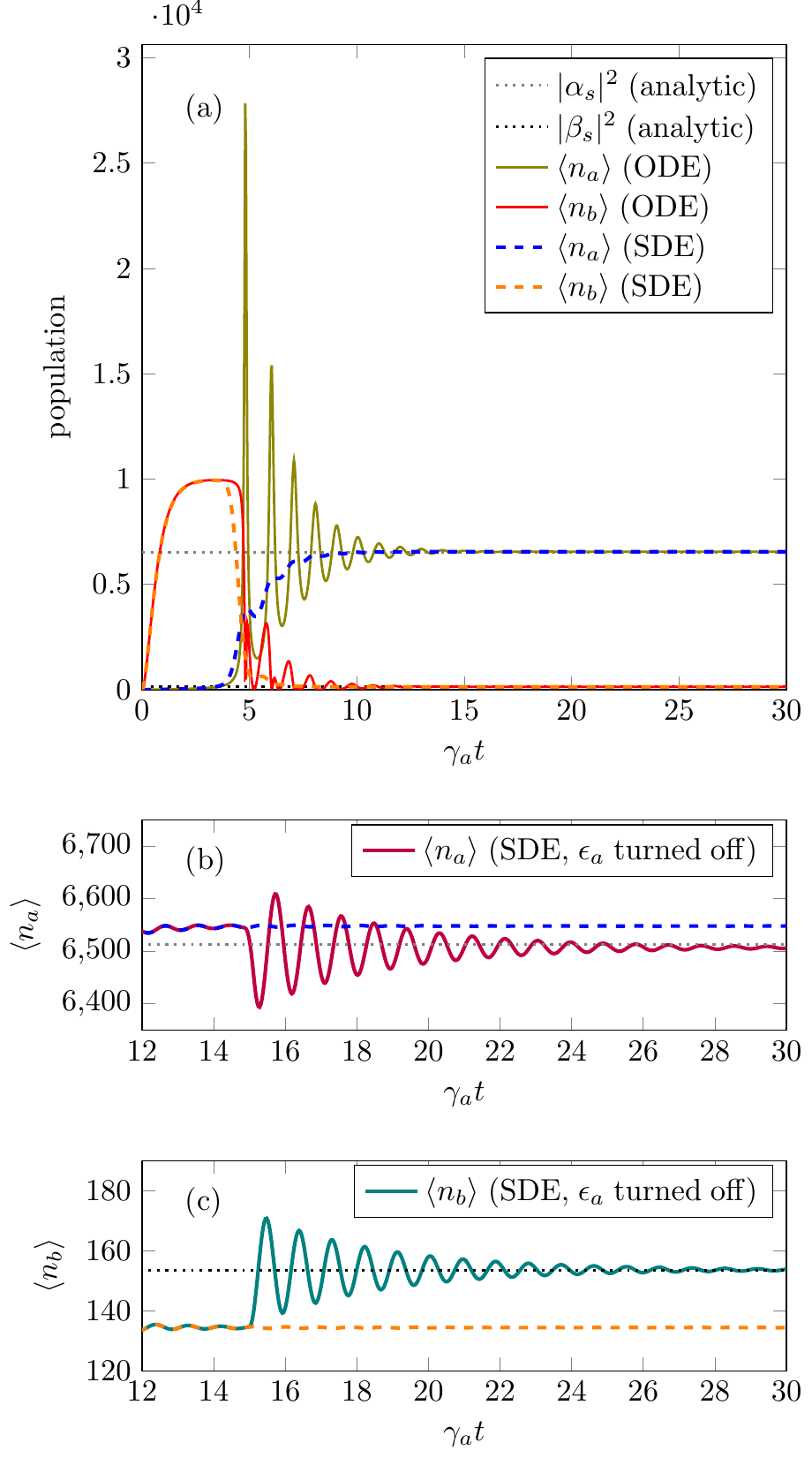}\\
	\caption{The time evolution of the expected mode populations for a system with the parameters $\kap=0.001\gam_a, \gam_b=2\gam_a, \epsilon_a = 5\gam_a$, and $\epsilon_b=200\gam_a$. Plot (a) compares simulations of the truncated positive-\textit{P} SDE \eqref{eq:ito} and the semi-classical ODE \eqref{eq:ode}. Plots (b) and (c) compare the truncated positive-\textit{P} simulation from (a) (dashed lines) with another simulation of \eqref{eq:ito} in which the injected signal $\epsilon_a$ is turned off at $\gam_at=15$ (continuous lines). Thin dotted lines indicate the analytically-determined semi-classical steady-state of \eqref{eq:ode}. Stochastic simulations were averaged over $10^6$ trajectories.}
	\label{fig:timeevolution}
\end{figure}

Here we introduce a weak injected signal $\epsilon_a\ll\epsilon_b$ to the equation for $\alpha$ (and $\epsilon_a^*$ for $\alpha^{\!+}$). This approach is also likely to be used in experiments as the low efficiency of the process all but prohibits triplet production, and it remains unclear whether continuous down conversion is even possible without a seed.
Using this aproach, the expected mode populations as given by \eqref{eq:ito} and \eqref{eq:ode} are compared in \fref{fig:timeevolution}(a). Inclusion of noise significantly reduces the magnitude of the oscillatory behaviour in the transient regime, but both models eventually reach the same steady-state. Removing the injected signal once the steady-state is reached, as shown in \fref{fig:timeevolution}(b) and (c), leads to small transients and a slightly altered steady-state. This steady-state is consistent with semi-classical values of $|\alpha_s|^2$ and $|\beta_s|^2$, determined analytically from \eqref{eq:ode_ss_a} and \eqref{eq:ode_ss_b}.

\begin{figure}
	\centering
	\includegraphics[width=\linewidth]{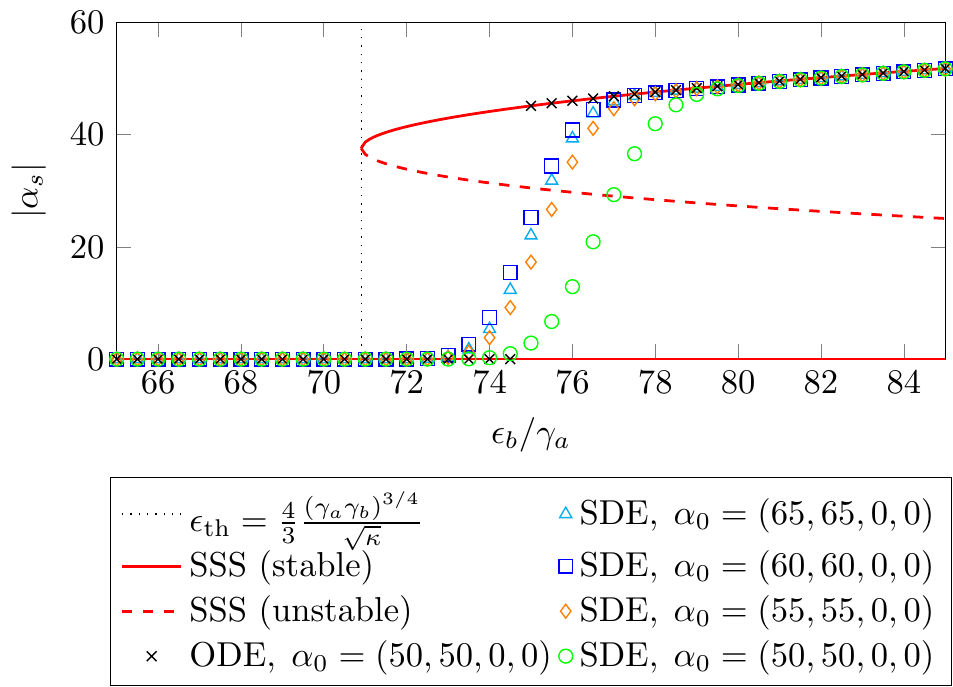}
	\caption{The steady-states of $|\alpha|$ as a function of pump intensity, as given by a steady-state solver (SSS) acting on the semi-classical ODE \eqref{eq:ode}, simulations of the truncated positive-\textit{P} SDE \eqref{eq:ito} with various initial states, and a simulation of the semi-classical ODE \eqref{eq:ode}. The parameters used are $\kap=0.001\gam_a$, $\gam_b=2\gam_a$, and $\epsilon_a=0$ (i.e. no injected signal). The non-vacuum initial states used were not chosen on any particular physical basis, but rather such that down conversion would proceed in the simulations. Simulations were run until $\gam_at=30$, and stochastic simulations averaged over $10^6$ trajectories.}
	\label{fig:endvalsunpumpeda}
\end{figure}

\begin{figure}
	\centering
	\includegraphics[width=\linewidth]{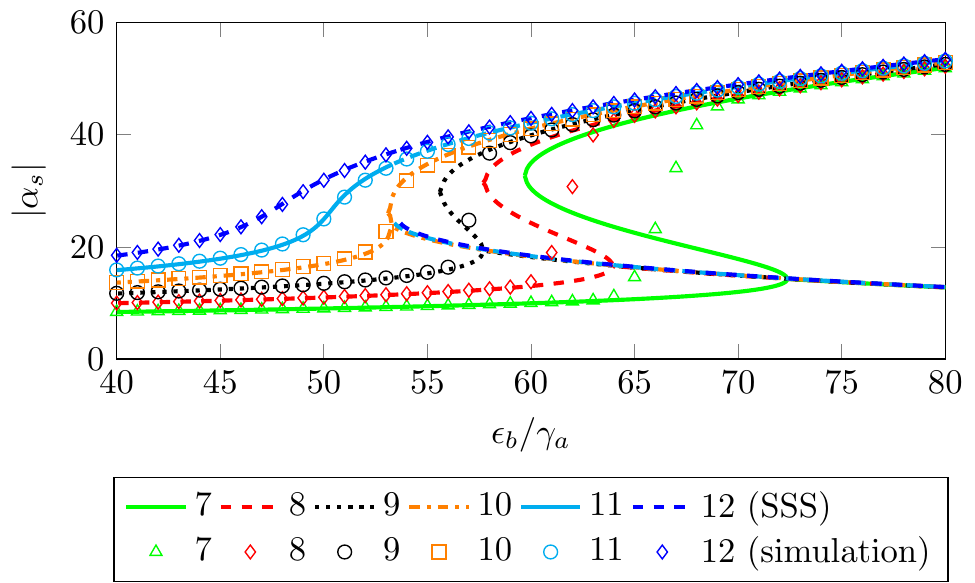}
	\caption{The steady-states of $|\alpha|$ as a function of the pump intensity $\epsilon_b/\gam_a$, for various values of the intensity of the injected signal $\epsilon_a$. A comparison is made between semi-classical results from a steady-state solver (lines) and simulations of the truncated positive-\textit{P} SDE \eqref{eq:ito} (points). The parameters used are $\kap=0.001\gam_a$, and $\gam_b=2\gam_a$, with the non-vacuum initial state taken to be $\bm{\alpha}=(30,30,0,0)$. This initial state was not chosen on any physical basis. Simulations were run until $\gam_at=100$, and averaged over $10^6$ trajectories.}
	\label{fig:endvalspumpeda}
\end{figure}

The time evolution and eventual steady-state presented in \fref{fig:timeevolution} corresponds to a particular choice of the pump intensity $\epsilon_b$. The corresponding steady-states of $|\alpha|$, as given by \eqref{eq:ito} and by \eqref{eq:ode}, are compared in \fref{fig:endvalsunpumpeda} as a function of the pump intensity $\epsilon_b$. The analytic steady-states of \eqref{eq:ode}, discussed above for $\gam_a\ll\gam_b$, are replicated in \fref{fig:endvalsunpumpeda} (red lines) for $\gam_a \not\ll\gam_b$, using the numeric steady-state solver from DifferentialEquations.jl \cite{DiffEq} acting on \eqref{eq:ode}. Direct simulation of \eqref{eq:ode} produces results consistent with the steady-state solver, but the transition from $\alpha=0$ to the upper branch will depend on the initial state and/or any injected signal. A single realisation is presented in \fref{fig:endvalsunpumpeda} (black crosses). The steady-states of $|\alpha|$ given by simulations of the truncated positive-\textit{P} SDE \eqref{eq:ito} are presented in \fref{fig:endvalsunpumpeda} for four different initial states (cyan triangles, blue squares, orange diamonds, green circles). As above, these simulations use an injected signal, $\epsilon_a$, which is turned off part way through the simulations. The results match the semi-classical steady-states for most pump intensities, except in the region immediately above the semi-classical pumping threshold, referred to henceforth as the transition region. In this region the steady-state populations take on values intermediate to, and form a connection between, the stable semi-classical solutions. Individual trajectories of \eqref{eq:ito} in the transition region end up at either $\alpha=0$, or stochastically fluctuating about the upper branch; the expectation values presented indicate the ratio of trajectories that go up or down, and are dependent on the initial state (as shown in \fref{fig:endvalsunpumpeda}) and/or the presence of any injected signal (see \fref{fig:endvalspumpeda}). It is important to emphasize that only operator moments can be recovered from these simulations, so the individual trajectories are not physically meaningful.

We now briefly consider the steady-states in the case that we do not turn off the injected signal, which has so far only been used to initiate down conversion. \fref{fig:endvalspumpeda} presents the steady-states around the transition region for systems with a range of injected signal intensities. Values given by simulations of \eqref{eq:ito} are presented alongside the (stable and unstable) semi-classical results of \eqref{eq:ode}, obtained using a numeric steady-state solver. For large injected signals there is only a single stable semi-classical solution at any given pump intensity. As the injected signal is decreased below some critical intensity, this solution morphs into what will eventually be the upper branch, the (unstable) lower branch, and part of the $\alpha=0$ solution. The other part of the $\alpha=0$ solution is in fact a separate solution which is only realized below the critical injected signal. Simulations of \eqref{eq:ito} are largely in agreement with these semi-classical results. Below the critical injected signal we observe the same behaviour in the transition region as in the $\epsilon_a=0$ results from \fref{fig:endvalsunpumpeda}, i.e. steady-states intermediate to the two stable semi-classical solutions. We emphasize that the steady-states will depend on the initial state of the system. In particular, steady-states intermediate to the two semi-classically stable solutions will not always exist; parameters were specifically chosen for \fref{fig:endvalspumpeda} to show that they are possible.

\section{Squeezing and entanglement}\label{sqz}
The quantities of interest in quantum information applications are usually the steady-state fluctuations of the output fields. It is standard quantum optical practice to consider the steady-state spectra, which are calculated via Fourier transform of the two-time covariance matrix. This approach is outlined in \cite{StochasticMethods}, having been originally developed for stochastic analysis of chemical reactions by Chaturvedi \textit{et al.} \cite{Gardiner1977}. We begin by decomposing the phase space variables into their steady-state expectation values and a fluctuation term,
\begin{align} \label{eq:ssDecompositon}
\bm{\alpha}(t) = \overline{\bm{\alpha}}_s + \delta\bm{\alpha}(t).
\end{align}
Substituting \eqref{eq:ssDecompositon} into \eqref{eq:ito} and linearising, the resulting equation of motion is the multivariate Ornstein--Uhlenbeck process
\begin{align} \label{eq:Ornstein-Uhlenbeck}
d\ \delta\bm{\alpha} = -A\delta\bm{\alpha} \ dt + B \ d\mathbf{W},
\end{align}
with
\begin{align}
&A =
\begin{bmatrix}
\gam_a                  & -2\kap\alpha_{\!s}^*\beta_s & -\kap\alpha_{\!s}^{*2} & 0                 \\
-2\kap\alpha_{\!s}\beta_s^* & \gam_a                  & 0                   & -\kap\alpha_{\!s}^2 \\
\kap\alpha_{\!s}^2          & 0                         & \gam_b            & 0                 \\
0                         & \kap\alpha_{\!s}^{*2}       & 0                   & \gam_b
\end{bmatrix}, \label{eq:DriftMatrix}
\end{align}
and
\begin{align}
&BB^T = D =
\begin{bmatrix}
2\kap\alpha_{\!s}^*\beta_{\!s}               & 0 & 0 & 0                 \\
0 & 2\kap\alpha_{\!s}\beta_{\!s}^*                  & 0                   & 0 \\
0          & 0                         & 0            & 0                 \\
0                         & 0      & 0                   & 0
\end{bmatrix}.
\label{eq:DiffusionMatrix}
\end{align}
The process of linearization assumes the fluctuations are small relative to their respective expectation values, which holds for all pump intensities except for those in the transition region. This region is examined in more detail in Appendix \ref{sec:fluctuations}. The Ornstein--Uhlenbeck processes \eqref{eq:Ornstein-Uhlenbeck} is stationary, so we can define the two-time covariance matrix as a function of a single variable,
\begin{align}
G(\tau) = \av{\delta\bm{\alpha}(t+\tau) \ \delta\bm{\alpha}^{\!\dagger}(t)},
\end{align}
and in turn define the Spectrum matrix,
\begin{align}
S(\omega) = \frac{1}{2\pi} \int_{-\infty}^{\infty} e^{-i\omega\tau} G(\tau) \ d\tau,
\end{align}
where $\omega$ is an angular frequency. Using standard results~\cite{StochasticMethods}, we can calculate the Spectrum matrix directly from \eqref{eq:DriftMatrix} and \eqref{eq:DiffusionMatrix} via
\begin{align} \label{eq:spectrum}
S(\omega) = (A+i\omega I)^{-1} D \, (A^\dagger-i\omega I)^{-1}\!,
\end{align}
where $I$ is the identity matrix. It is important to note that this process is not valid if the eigenvalues of $A$ have any negative real parts, as this will result in fluctuations increasing without bound. However, this is not the case for any of the results presented.

In order to express the spectrum in terms of the canonical quadrature operators $\hat{X}_a = \hat{a}^\dagger + \hat{a}, \hat{Y}_a=i(\hat{a}^\dagger - \hat{a})$, and those defined analogously for the $b$ mode, we calculate \cite{WallsMilburn}
\begin{align}
S^q(\omega) = QS(\omega)Q^T, \qquad \text{where} \quad Q=
\begin{bmatrix}
1  & 1 & 0  & 0 \\
-i & i & 0  & 0 \\
0  & 0 & 1  & 1 \\
0  & 0 & -i & i
\end{bmatrix}.
\end{align}
The output spectra is calculated using the standard input--output relations \cite{InputOutput}, hence we can calculate output spectral variances and covariances via
\begin{align} \label{eq:covarianceSpectra}
\begin{split}
\mathrm{C}[\hat{X}_i,\hat{X}_j](\omega)  &= \delta_{ij} + \sqrt{\gam_i\gam_j} (S^q_{2i-1,2j-1} + S^q_{2j-1,2i-1}),\\
\mathrm{C}[\hat{Y}_i,\hat{Y}_j](\omega)  &= \delta_{ij} + \sqrt{\gam_i\gam_j} (S^q_{2i,2j} + S^q_{2j,2i}),
\end{split}
\end{align}
where we use $a\rightarrow1$ and $b\rightarrow2$ for the $i,j$ indexing, required to be cyclic in all indices.

\begin{figure}
	\centering
	\includegraphics[width=0.9\linewidth]{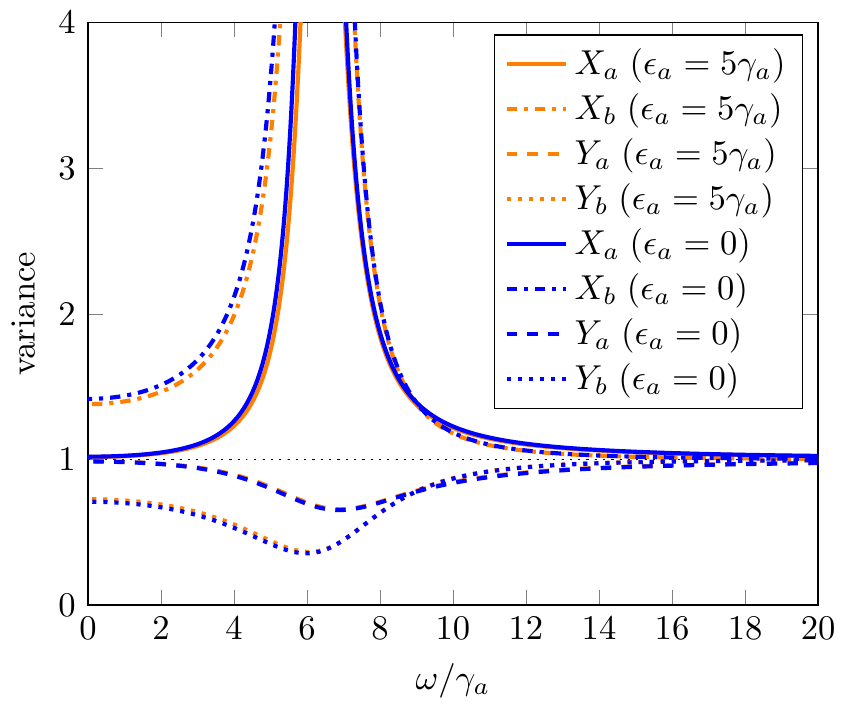}
	\caption{The spectrum of steady-state quadrature variances of the output fields for both modes in the cases that an injected signal is (orange) and is not (blue) present in the steady-state. The $\hat{Y}_a$ and $\hat{Y}_b$ quadratures have variances less than that of the vacuum, and hence both modes are squeezed. The parameters used are $\kap=0.001\gam_a, \gam_b=2\gam_a, \epsilon_a = 5\gam_a$ or 0, and $\epsilon_b=200\gam_a$, and simulations were averaged over $10^6$ trajectories.}
	\label{fig:squeezingcomparison}
\end{figure}

\begin{figure}
	\centering
	\includegraphics[width=\linewidth]{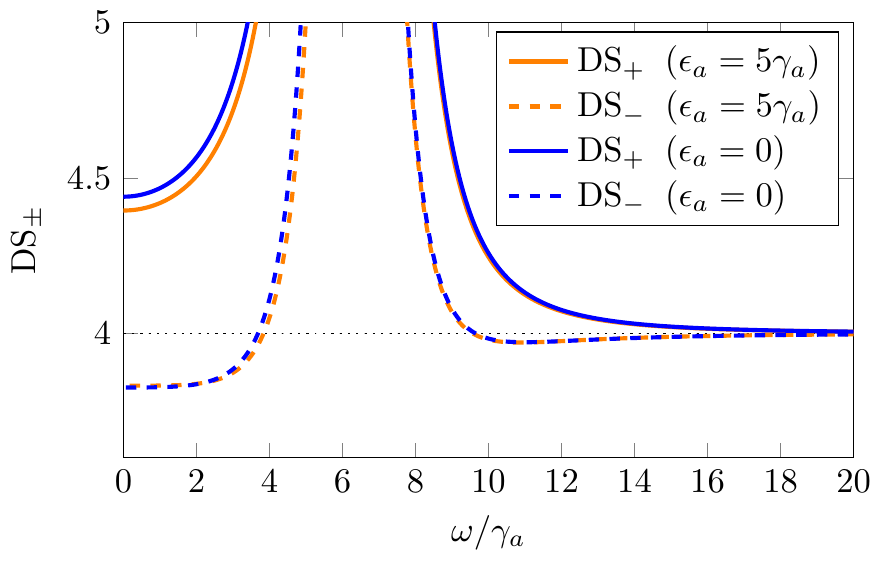}
	\caption{The steady-state Duan--Simon spectra of the output fields for the cases that an injected signal is (orange) and is not (blue) present in the steady-state. The Duan--Simon inequality \eqref{eq:DS} is violated by $\mathrm{DS}_-$, indicating that the two fields are entangled. The parameters used are $\kap=0.001\gam_a, \gam_b=2\gam_a, \epsilon_a = 5\gam_a$ or 0, and $\epsilon_b=200\gam_a$, and simulations were averaged over $10^6$ trajectories.}
	\label{fig:dscomparison}
\end{figure}

The non-classical nature of the output fields can be demonstrated in various ways. One of the common measures is whether the uncertainty in one quadrature is less than the vacuum (or coherent state) level, in which case the output is said to be squeezed \cite{Caves, walls_squeezed_1983}. In the present system we find squeezing of the $\hat{Y}$ quadratures, shown in \fref{fig:squeezingcomparison} for the simulation presented in \fref{fig:timeevolution}(a). Also of interest is entanglement and inseperability of the two output fields, often demonstrated by violation of the Duan--Simon inequality \cite{Duan,Simon}, which can be written as
\begin{align} \label{eq:DS}
\mathrm{DS}_\pm \equiv \mathrm{V}(\hat{X}_i \pm \hat{X}_j) + \mathrm{V}(\hat{Y}_i \mp \hat{Y}_j) \geq 4.
\end{align}
We find that the modes are entangled, with $D_-$ clearly violated for $\omega\rightarrow0$, as shown in \fref{fig:dscomparison}. Violation of the $\mathrm{DS}_-$ inequality is also predicted at higher frequencies, but at a magnitude that is unlikely to be experimentally measurable. It should be noted that we are dealing with a non-Gaussian system, so in regions where \eqref{eq:DS} is satisfied we cannot necessarily claim that the fields are not entangled.

\begin{figure}
	\centering\
	\subfigure{\includegraphics[width=\linewidth]{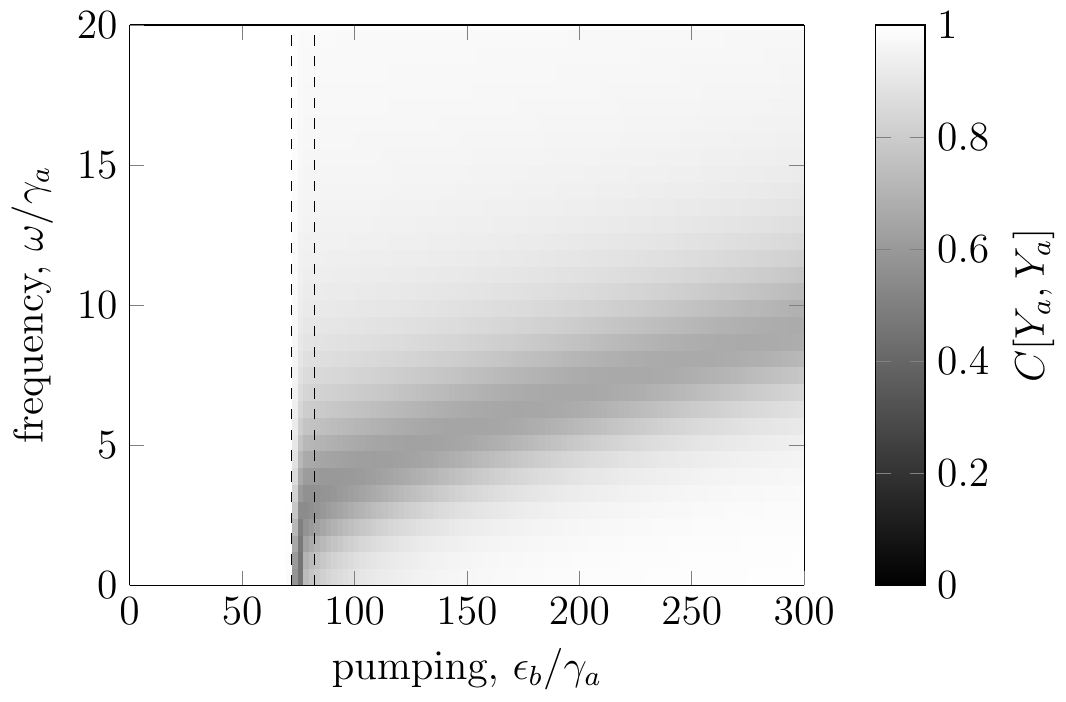}} \\
	\subfigure{\includegraphics[width=\linewidth]{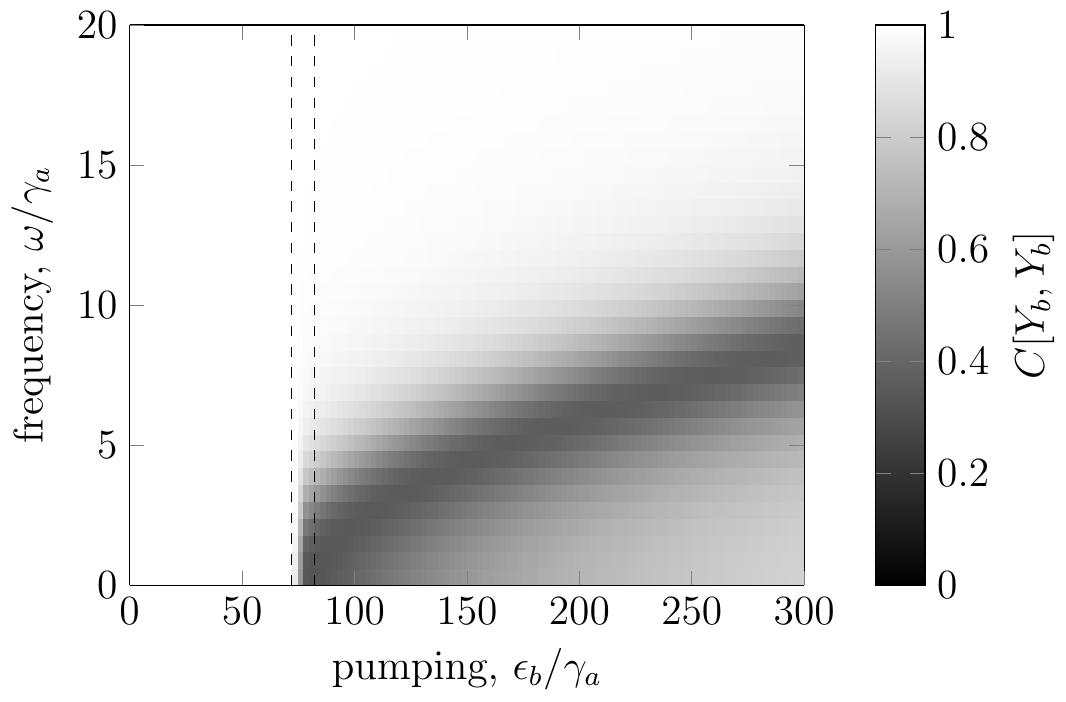}}\
	\caption{The spectrum of quadrature variances for the output fields as a function of the pump intensity $\epsilon_b$. The parameters used are $\kap=0.001\gam_a, \gam_b=2\gam_a$ and $\epsilon_a=0$, and simulations were averaged over $10^6$ trajectories. The region enclosed by the dashed lines indicates where the linearized fluctuation analysis does not apply due to the magnitude of fluctuations - see Appendix \ref{sec:fluctuations} for details.}
	\label{fig:vy}
\end{figure}

The steady-states obtained from the simulations examined in \fref{fig:endvalsunpumpeda} were used to investigate squeezing and bipartite entanglement over a range of pump intensities. The linearized fluctuation analysis applied here cannot be applied to the transition region due to the magnitude of the fluctuations. The squeezing results are presented in \fref{fig:vy}. Below threshold the low-energy mode is in a vacuum state, and the classically pumped high energy mode is in a coherent state, so the uncertainty is at the vacuum level. Above threshold we observe squeezing at all pump intensities up to $\epsilon_b/\gam_a=300$, the largest value examined. The spectra have the qualitative shape of those presented in \fref{fig:squeezingcomparison}. As the pump intensity increases, the maximal squeezing moves towards higher frequencies and decreases in magnitude. The key difference between the high- and low-energy modes is that near the transition region the high energy ($b$) mode exhibits maximal squeezing at zero-frequency. The $b$ mode also exhibits stronger squeezing at all pump intensities, and the magnitude of its squeezing decreases slower than that of the low energy mode as the pump intensity increases.

We find entanglement of the modes at all pump intensities above the transition region up to at least $\epsilon_b/\gam_a=300$. The $DS_-$ inequality is always violated at large and small frequencies, but is satisfied between these two regions, just as in \fref{fig:dscomparison}. $DS_+$ is never violated. The maximal violation of the $\mathrm{DS}_-$ inequality decreases in magnitude as the pumping is increased, indicating less entanglement. The peak in $DS_-$ [$DS_+$] moves towards larger frequencies and increases [decreases] in magnitude as the pumping is increased.

\section{Discussion and Conclusions}\label{disc}
\subsection{Discussion}
The requirement to seed the process at $\omega_a$ is a significant experimental challenge.
The seed must not only be as close as possible to satisfying $\omega_a = \omega_b/3$, but it must also be phase stable and have the right phase relative to the pump laser; a phase shift of $\pi/3$ causes the gain in the non-trivial solution to turn negative and the signal will decay, similar to the behavior of a degenerate second-order optical parametric amplifier with a pump phase shift of $\pi/2$.
This challenge can be mitigated by only seeding the interaction for a time long enough to move the system close to steady state, but short enough that the relative phase varies by less that $\pi/3$. It is worth noting that relative phase stability of the seed and pump does not pose a problem for non-degenerate third-order down conversion (not considered here), as phase wandering is compensated by complementary wandering in the unseeded modes. Although tripartite entanglement can be created via tuned pair downconversion~\cite{BrightTripartite}, the non-degenerate system could also be useful as a more direct source of tripartite quantum correlations.

An alternative approach is to derive the pump from the third harmonic of the seed laser. This automatically phase locks the system provided the optical path length is stable, and additionally has the benefit that the seed laser can act as the local oscillator for an optical homodyne measurement.
If the pump were generated by third-harmonic generation, however, it could be amplitude squeezed, and it is unclear how this would impact the phase squeezing created by triplet down conversion.

Our choice of relatively large nonlinearity $\kappa = 0.001\gamma_a$ enabled a study of significant nonlinearity in the system, including the intermediate steady states observed in the transition region. In the context of making a comparison between truncated positive-$P$ and the number state basis (see Appendix \ref{sec:MCWF}) our choice of $\kappa$ acts to over-estimate the discrepancy between the two methods, amplifying the role of nonlinearity relative to dissipation, compared to more readily accessible experimental parameters (see Sec.~\ref{expc}). Despite this choice, we observe good agreement between the truncated positive-$P$ and Monte Carlo wavefunction methods.
\subsection{Conclusion}
We have performed a numerical investigation of intracavity degenerate triplet down conversion using a truncated positive-\textit{P} representation. A systematic analysis of the stable steady-states of the expected mode populations found close agreement with semi-classical predictions, except in the region immediately above the pumping threshold. Interestingly, here we find that quantum stochastic simulations predict the possibility of steady states intermediate to the two stable semi-classical solutions.

A steady-state linearized fluctuation analysis was performed and used to calculate the intracavity fluctuation spectrum. From this, regimes of squeezing and bipartite entanglement were identified. Both of these quantities were found to be maximal closest to the pumping threshold. However, the linearisation involved in determining the spectrum is not valid immediately above the pumping threshold. An interesting future direction would be to carry out a systematic study of critical fluctuations, and their measurable extracavity signatures.

\acknowledgements
AB and HS are supported by the Dodd-Walls Centre for Photonic and Quantum Technologies.

\appendix
\section{Validity of Linearized Fluctuation Analysis} \label{sec:fluctuations}
The Ornstein--Uhlenbeck process \eqref{eq:Ornstein-Uhlenbeck} used to determine the spectra presented in this work was derived by linearising the steady-state equations of motion under the assumption that fluctuations of the phase-space variables are negligible compared to their expectation values. This assumption is examined by considering the quadrature operators for both modes, rather than the phase-space variables themselves, because they are explicitly real-valued quantities. However, this introduces the question as to which quantities to compare to which. For the systems examined in this paper, the steady-state (quadrature) phase-space distributions for each mode are centred on the $X$ axis (i.e. $\av{\hat{Y}}=0$), so we must consider the standard deviation of both the $X$ and $Y$ quadratures relative to $\av{\hat{X}}$. These values are calculated by averaging over stochastic trajectories. For example,
\begin{align}
\overline{\alpha+\alpha^{\!+}} &\rightarrow \av{a+a^\dagger} = \av{X_a},\\
\overline{1+2\alpha\alpha^{\!+}+\alpha^2 +\alpha^{\!+2}} &\rightarrow \av{(\hat{a}+\hat{a}^\dagger)^2} = \av{\hat{X}_a^2},
\end{align}
and then $\Delta X_a = \sqrt{\av{\hat{X}_a^2} - \av{\hat{X}_a}^2}$.

\begin{figure}[!t]
	\centering
	\includegraphics[width=0.9\linewidth]{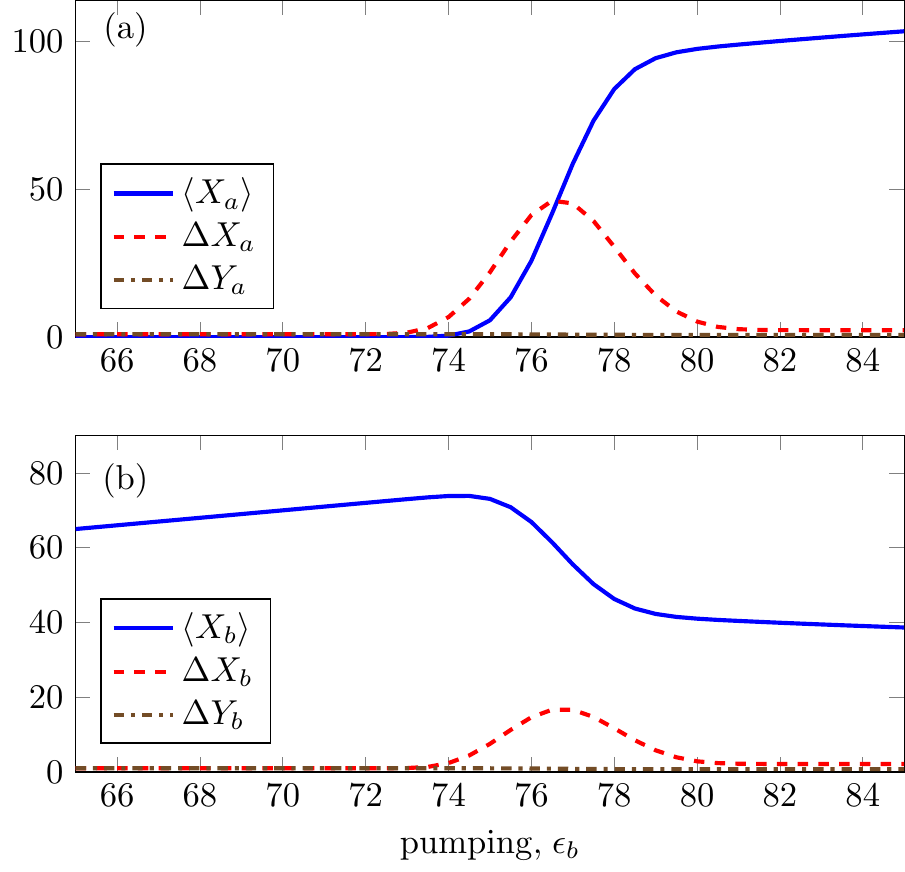}
	\caption{For each mode, the steady-state standard deviations of the $X$ and $Y$ quadratures are compared to expectation value of the $X$ quadrature over a range of pump intensities, including the transition region. Fluctuations are significant in the transition region, but are otherwise negligible. The parameters used are $\kap=0.001\gam_a, \gam_b=2\gam_a$ and $\epsilon_a=0$, and simulations were averaged over $10^6$ trajectories.}
	\label{fig:fluctuations}
\end{figure}

The results are presented over a range of pump intensities in \fref{fig:fluctuations}(a) for the low energy mode, and (b) for the high energy mode. The quadrature fluctuations exhibits the same behaviour in both modes. While $\Delta Y$ is negligible everywhere, $\Delta X$ becomes comparable to $\av{\hat{X}}$ within the transition region. In context of the phase-space variables, this corresponds to some trajectories going towards each of the two distinct stable semi-classical solutions, resulting in a large range of values. For this reason, we note that our method for determining the spectra is not valid within the transition region. It should be noted that just like the expectation value, the standard deviation is dependent on the initial conditions. While the results presented here correspond to a single set of parameters, they are representative of the general behavior of the system.

\section{Comparison of Truncated Positive-\textit{P} and Monte Carlo Wave Function Techniques} \label{sec:MCWF}
The truncation involved in the mapping from \eref{eq:FPE} to \eref{eq:ito} inevitably introduces a level of discrepancy, but will generally be more valid at higher intensities due to the scaling of the third-order terms with respect to the number of photons. This claim is examined via direct simulation of \eqref{eq:master} using a Monte Carlo wave function (MCWF) technique \cite{MCWF-Dalibard,MCWF-Dum,DaleyReview} in a number state basis.

\begin{figure}[!t]
 	\centering
 	\includegraphics[width=\linewidth]{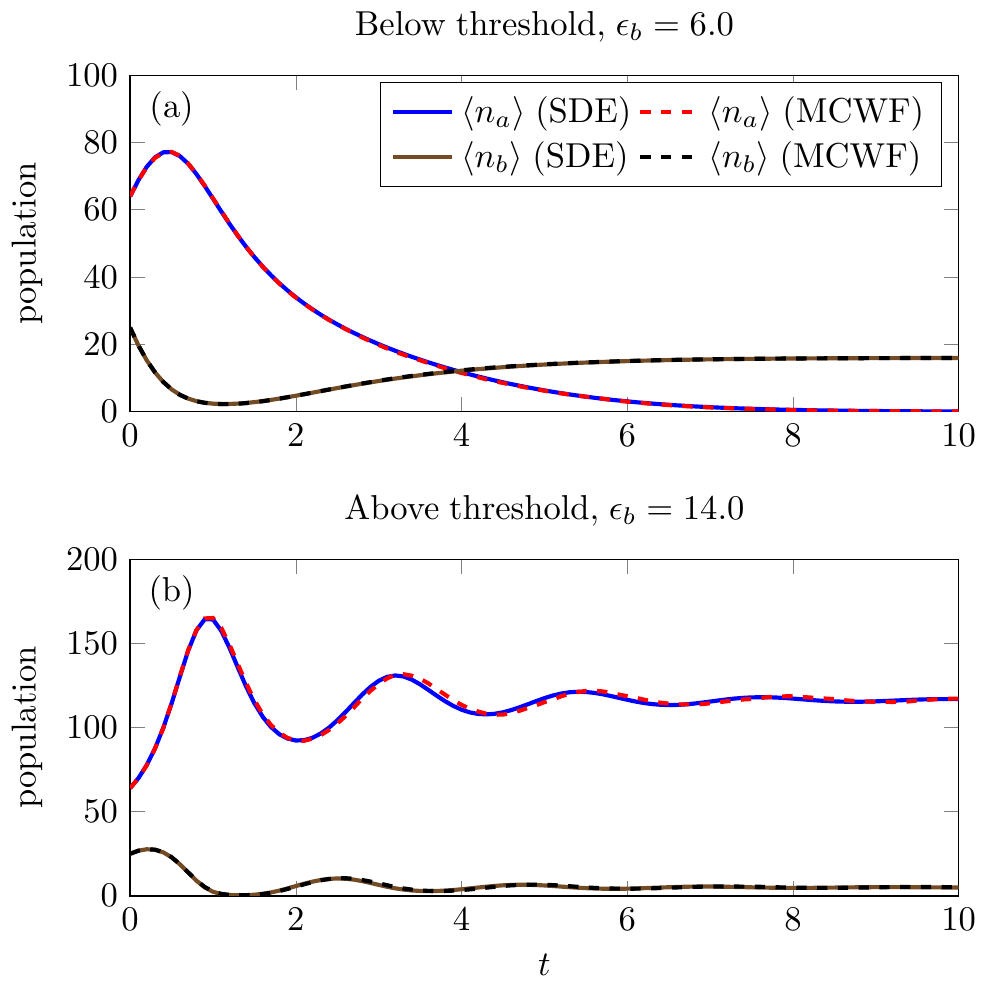}
 	\caption{A comparison between the time evolution of the expected mode populations, as given by the truncated positive-\textit{P} SDE \eqref{eq:ito} and by the MCWF technique, for systems both above and below threshold. The parameters used are $\kap=0.025, \gam_a=0.6, \gam_b=1.5$ and $\epsilon_a=0$, and the SDE [MCWF] was averaged over $10^6$ [1,000] trajectories. The MCWF simulations used Hilbert spaces with the dimensions $N_a=270$ and $N_b=90$ for the respective modes.}
 	\label{fig:MCWF}
 \end{figure}

 These simulations were implemented using the QuantumOptics.jl package \cite{QO}. Optical phase space techniques are used in these kinds of systems specifically because they scale well with system size, while a number state basis does not. For this reason it is not feasible to run MCWF simulations with the parameters used in this paper. Instead, we compare MCWF and truncated positive-\textit{P} simulations for a smaller system of $\sim\!100$ photons. The two techniques are compared in \fref{fig:MCWF}. Both above and below threshold the two techniques are in close agreement, although there is a slight difference in the period of the oscillations exhibited in (b).

For this reason, the effect of the truncation involved in deriving the SDEs used throughout this paper can be taken to be negligible. Both above and below threshold the MCWF simulations used Hilbert spaces with the dimensions $N_a=270$ and $N_b=90$ for the respective modes. It has been confirmed that increasing the size of these Hilbert spaces does not change the result.

\bibliographystyle{apsrev4-1}

\begin{thebibliography}{42}%
\makeatletter
\providecommand \@ifxundefined [1]{%
 \@ifx{#1\undefined}
}%
\providecommand \@ifnum [1]{%
 \ifnum #1\expandafter \@firstoftwo
 \else \expandafter \@secondoftwo
 \fi
}%
\providecommand \@ifx [1]{%
 \ifx #1\expandafter \@firstoftwo
 \else \expandafter \@secondoftwo
 \fi
}%
\providecommand \natexlab [1]{#1}%
\providecommand \enquote  [1]{``#1''}%
\providecommand \bibnamefont  [1]{#1}%
\providecommand \bibfnamefont [1]{#1}%
\providecommand \citenamefont [1]{#1}%
\providecommand \href@noop [0]{\@secondoftwo}%
\providecommand \href [0]{\begingroup \@sanitize@url \@href}%
\providecommand \@href[1]{\@@startlink{#1}\@@href}%
\providecommand \@@href[1]{\endgroup#1\@@endlink}%
\providecommand \@sanitize@url [0]{\catcode `\\12\catcode `\$12\catcode
  `\&12\catcode `\#12\catcode `\^12\catcode `\_12\catcode `\%12\relax}%
\providecommand \@@startlink[1]{}%
\providecommand \@@endlink[0]{}%
\providecommand \url  [0]{\begingroup\@sanitize@url \@url }%
\providecommand \@url [1]{\endgroup\@href {#1}{\urlprefix }}%
\providecommand \urlprefix  [0]{URL }%
\providecommand \Eprint [0]{\href }%
\providecommand \doibase [0]{http://dx.doi.org/}%
\providecommand \selectlanguage [0]{\@gobble}%
\providecommand \bibinfo  [0]{\@secondoftwo}%
\providecommand \bibfield  [0]{\@secondoftwo}%
\providecommand \translation [1]{[#1]}%
\providecommand \BibitemOpen [0]{}%
\providecommand \bibitemStop [0]{}%
\providecommand \bibitemNoStop [0]{.\EOS\space}%
\providecommand \EOS [0]{\spacefactor3000\relax}%
\providecommand \BibitemShut  [1]{\csname bibitem#1\endcsname}%
\let\auto@bib@innerbib\@empty
\bibitem [{\citenamefont {Olsen}(2017)}]{Olsen_2017}%
  \BibitemOpen
  \bibfield  {author} {\bibinfo {author} {\bibfnamefont {M.~K.}\ \bibnamefont
  {Olsen}},\ }\href {\doibase 10.1103/PhysRevA.96.063839} {\bibfield  {journal}
  {\bibinfo  {journal} {Phys. Rev. A}\ }\textbf {\bibinfo {volume} {96}},\
  \bibinfo {pages} {063839} (\bibinfo {year} {2017})}\BibitemShut {NoStop}%
\bibitem [{\citenamefont {Kolker}\ \emph {et~al.}(2018)\citenamefont {Kolker},
  \citenamefont {Kostyukova}, \citenamefont {Boyko}, \citenamefont {Badikov},
  \citenamefont {Badikov}, \citenamefont {Shadrintseva}, \citenamefont
  {Tretyakova}, \citenamefont {Zenov}, \citenamefont {Karapuzikov},\ and\
  \citenamefont {Zondy}}]{kolker_widely_2018}%
  \BibitemOpen
  \bibfield  {author} {\bibinfo {author} {\bibfnamefont {D.~B.}\ \bibnamefont
  {Kolker}}, \bibinfo {author} {\bibfnamefont {N.~Y.}\ \bibnamefont
  {Kostyukova}}, \bibinfo {author} {\bibfnamefont {A.~A.}\ \bibnamefont
  {Boyko}}, \bibinfo {author} {\bibfnamefont {V.~V.}\ \bibnamefont {Badikov}},
  \bibinfo {author} {\bibfnamefont {D.~V.}\ \bibnamefont {Badikov}}, \bibinfo
  {author} {\bibfnamefont {A.~G.}\ \bibnamefont {Shadrintseva}}, \bibinfo
  {author} {\bibfnamefont {N.~N.}\ \bibnamefont {Tretyakova}}, \bibinfo
  {author} {\bibfnamefont {K.~G.}\ \bibnamefont {Zenov}}, \bibinfo {author}
  {\bibfnamefont {A.~A.}\ \bibnamefont {Karapuzikov}}, \ and\ \bibinfo {author}
  {\bibfnamefont {J.-J.}\ \bibnamefont {Zondy}},\ }\href {\doibase
  10.1088/2399-6528/aab007} {\bibfield  {journal} {\bibinfo  {journal} {J.
  Phys. Comm.}\ }\textbf {\bibinfo {volume} {2}},\ \bibinfo {pages} {035039}
  (\bibinfo {year} {2018})}\BibitemShut {NoStop}%
\bibitem [{\citenamefont {Olsen}(2018)}]{Olsen:2018gr}%
  \BibitemOpen
  \bibfield  {author} {\bibinfo {author} {\bibfnamefont {M.~K.}\ \bibnamefont
  {Olsen}},\ }\href {\doibase 10.1016/j.optcom.2017.09.090} {\bibfield
  {journal} {\bibinfo  {journal} {Opt. Comm.}\ }\textbf {\bibinfo {volume}
  {410}},\ \bibinfo {pages} {966} (\bibinfo {year} {2018})}\BibitemShut
  {NoStop}%
\bibitem [{\citenamefont {Li}\ and\ \citenamefont {Olsen}(2018)}]{Li_2018}%
  \BibitemOpen
  \bibfield  {author} {\bibinfo {author} {\bibfnamefont {J.}~\bibnamefont
  {Li}}\ and\ \bibinfo {author} {\bibfnamefont {M.~K.}\ \bibnamefont {Olsen}},\
  }\href {\doibase 10.1103/PhysRevA.97.043856} {\bibfield  {journal} {\bibinfo
  {journal} {\pra}\ }\textbf {\bibinfo {volume} {97}},\ \bibinfo {pages}
  {043856} (\bibinfo {year} {2018})}\BibitemShut {NoStop}%
\bibitem [{\citenamefont {Fürst}\ \emph {et~al.}(2011)\citenamefont {Fürst},
  \citenamefont {Strekalov}, \citenamefont {Elser}, \citenamefont {Aiello},
  \citenamefont {Andersen}, \citenamefont {Marquardt},\ and\ \citenamefont
  {Leuchs}}]{furst_quantum_2011}%
  \BibitemOpen
  \bibfield  {author} {\bibinfo {author} {\bibfnamefont {J.~U.}\ \bibnamefont
  {Fürst}}, \bibinfo {author} {\bibfnamefont {D.~V.}\ \bibnamefont
  {Strekalov}}, \bibinfo {author} {\bibfnamefont {D.}~\bibnamefont {Elser}},
  \bibinfo {author} {\bibfnamefont {A.}~\bibnamefont {Aiello}}, \bibinfo
  {author} {\bibfnamefont {U.~L.}\ \bibnamefont {Andersen}}, \bibinfo {author}
  {\bibfnamefont {C.}~\bibnamefont {Marquardt}}, \ and\ \bibinfo {author}
  {\bibfnamefont {G.}~\bibnamefont {Leuchs}},\ }\href {\doibase
  10.1103/PhysRevLett.106.113901} {\bibfield  {journal} {\bibinfo  {journal}
  {\prl}\ }\textbf {\bibinfo {volume} {106}},\ \bibinfo {pages} {113901}
  (\bibinfo {year} {2011})}\BibitemShut {NoStop}%
\bibitem [{\citenamefont {Peano}\ \emph {et~al.}(2015)\citenamefont {Peano},
  \citenamefont {Schwefel}, \citenamefont {Marquardt},\ and\ \citenamefont
  {Marquardt}}]{Peano_2015}%
  \BibitemOpen
  \bibfield  {author} {\bibinfo {author} {\bibfnamefont {V.}~\bibnamefont
  {Peano}}, \bibinfo {author} {\bibfnamefont {H.~G.~L.}\ \bibnamefont
  {Schwefel}}, \bibinfo {author} {\bibfnamefont {C.}~\bibnamefont {Marquardt}},
  \ and\ \bibinfo {author} {\bibfnamefont {F.}~\bibnamefont {Marquardt}},\
  }\href {\doibase 10.1103/PhysRevLett.115.243603} {\bibfield  {journal}
  {\bibinfo  {journal} {\prl}\ }\textbf {\bibinfo {volume} {115}},\ \bibinfo
  {pages} {243603} (\bibinfo {year} {2015})}\BibitemShut {NoStop}%
\bibitem [{\citenamefont {Andersen}\ \emph {et~al.}(2016)\citenamefont
  {Andersen}, \citenamefont {Gehring}, \citenamefont {Marquardt},\ and\
  \citenamefont {Leuchs}}]{andersen_30_2016}%
  \BibitemOpen
  \bibfield  {author} {\bibinfo {author} {\bibfnamefont {U.~L.}\ \bibnamefont
  {Andersen}}, \bibinfo {author} {\bibfnamefont {T.}~\bibnamefont {Gehring}},
  \bibinfo {author} {\bibfnamefont {C.}~\bibnamefont {Marquardt}}, \ and\
  \bibinfo {author} {\bibfnamefont {G.}~\bibnamefont {Leuchs}},\ }\href
  {\doibase 10.1088/0031-8949/91/5/053001} {\bibfield  {journal} {\bibinfo
  {journal} {Phys. Scr.}\ }\textbf {\bibinfo {volume} {91}},\ \bibinfo {pages}
  {053001} (\bibinfo {year} {2016})}\BibitemShut {NoStop}%
\bibitem [{\citenamefont {Vahlbruch}\ \emph {et~al.}(2016)\citenamefont
  {Vahlbruch}, \citenamefont {Mehmet}, \citenamefont {Danzmann},\ and\
  \citenamefont {Schnabel}}]{vahlbruch_detection_2016}%
  \BibitemOpen
  \bibfield  {author} {\bibinfo {author} {\bibfnamefont {H.}~\bibnamefont
  {Vahlbruch}}, \bibinfo {author} {\bibfnamefont {M.}~\bibnamefont {Mehmet}},
  \bibinfo {author} {\bibfnamefont {K.}~\bibnamefont {Danzmann}}, \ and\
  \bibinfo {author} {\bibfnamefont {R.}~\bibnamefont {Schnabel}},\ }\href
  {\doibase 10.1103/PhysRevLett.117.110801} {\bibfield  {journal} {\bibinfo
  {journal} {\prl}\ }\textbf {\bibinfo {volume} {117}},\ \bibinfo {pages}
  {110801} (\bibinfo {year} {2016})}\BibitemShut {NoStop}%
\bibitem [{\citenamefont {Takei}\ \emph {et~al.}(2005)\citenamefont {Takei},
  \citenamefont {Yonezawa}, \citenamefont {Aoki},\ and\ \citenamefont
  {Furusawa}}]{takei_high-fidelity_2005}%
  \BibitemOpen
  \bibfield  {author} {\bibinfo {author} {\bibfnamefont {N.}~\bibnamefont
  {Takei}}, \bibinfo {author} {\bibfnamefont {H.}~\bibnamefont {Yonezawa}},
  \bibinfo {author} {\bibfnamefont {T.}~\bibnamefont {Aoki}}, \ and\ \bibinfo
  {author} {\bibfnamefont {A.}~\bibnamefont {Furusawa}},\ }\href {\doibase
  10.1103/PhysRevLett.94.220502} {\bibfield  {journal} {\bibinfo  {journal}
  {\prl}\ }\textbf {\bibinfo {volume} {94}},\ \bibinfo {pages} {220502}
  (\bibinfo {year} {2005})}\BibitemShut {NoStop}%
\bibitem [{\citenamefont {Jing}\ \emph {et~al.}(2003)\citenamefont {Jing},
  \citenamefont {Zhang}, \citenamefont {Yan}, \citenamefont {Zhao},
  \citenamefont {Xie},\ and\ \citenamefont {Peng}}]{jing_experimental_2003}%
  \BibitemOpen
  \bibfield  {author} {\bibinfo {author} {\bibfnamefont {J.}~\bibnamefont
  {Jing}}, \bibinfo {author} {\bibfnamefont {J.}~\bibnamefont {Zhang}},
  \bibinfo {author} {\bibfnamefont {Y.}~\bibnamefont {Yan}}, \bibinfo {author}
  {\bibfnamefont {F.}~\bibnamefont {Zhao}}, \bibinfo {author} {\bibfnamefont
  {C.}~\bibnamefont {Xie}}, \ and\ \bibinfo {author} {\bibfnamefont
  {K.}~\bibnamefont {Peng}},\ }\href {\doibase 10.1103/PhysRevLett.90.167903}
  {\bibfield  {journal} {\bibinfo  {journal} {\prl}\ }\textbf {\bibinfo
  {volume} {90}},\ \bibinfo {pages} {167903} (\bibinfo {year}
  {2003})}\BibitemShut {NoStop}%
\bibitem [{\citenamefont {Cavanna}\ \emph {et~al.}(2016)\citenamefont
  {Cavanna}, \citenamefont {Just}, \citenamefont {Jiang}, \citenamefont
  {Leuchs}, \citenamefont {Chekhova}, \citenamefont {Russell},\ and\
  \citenamefont {Joly}}]{Cavanna:16}%
  \BibitemOpen
  \bibfield  {author} {\bibinfo {author} {\bibfnamefont {A.}~\bibnamefont
  {Cavanna}}, \bibinfo {author} {\bibfnamefont {F.}~\bibnamefont {Just}},
  \bibinfo {author} {\bibfnamefont {X.}~\bibnamefont {Jiang}}, \bibinfo
  {author} {\bibfnamefont {G.}~\bibnamefont {Leuchs}}, \bibinfo {author}
  {\bibfnamefont {M.~V.}\ \bibnamefont {Chekhova}}, \bibinfo {author}
  {\bibfnamefont {P.~S.}\ \bibnamefont {Russell}}, \ and\ \bibinfo {author}
  {\bibfnamefont {N.~Y.}\ \bibnamefont {Joly}},\ }\href {\doibase
  10.1364/OPTICA.3.000952} {\bibfield  {journal} {\bibinfo  {journal} {Optica}\
  }\textbf {\bibinfo {volume} {3}},\ \bibinfo {pages} {952} (\bibinfo {year}
  {2016})}\BibitemShut {NoStop}%
\bibitem [{\citenamefont {Drummond}\ and\ \citenamefont
  {Gardiner}(1980)}]{Drummond_1980}%
  \BibitemOpen
  \bibfield  {author} {\bibinfo {author} {\bibfnamefont {P.~D.}\ \bibnamefont
  {Drummond}}\ and\ \bibinfo {author} {\bibfnamefont {C.~W.}\ \bibnamefont
  {Gardiner}},\ }\href {\doibase 10.1088/0305-4470/13/7/018} {\bibfield
  {journal} {\bibinfo  {journal} {J. Phys. A}\ }\textbf {\bibinfo {volume}
  {13}},\ \bibinfo {pages} {2353} (\bibinfo {year} {1980})}\BibitemShut
  {NoStop}%
\bibitem [{\citenamefont {Bajer}(1991)}]{Bajer}%
  \BibitemOpen
  \bibfield  {author} {\bibinfo {author} {\bibfnamefont {J.}~\bibnamefont
  {Bajer}},\ }\href {\doibase 10.1080/09500349114551121} {\bibfield  {journal}
  {\bibinfo  {journal} {J. Mod. Opt.}\ }\textbf {\bibinfo {volume} {38}},\
  \bibinfo {pages} {1085} (\bibinfo {year} {1991})}\BibitemShut {NoStop}%
\bibitem [{\citenamefont {Felbinger}\ \emph {et~al.}(1998)\citenamefont
  {Felbinger}, \citenamefont {Schiller},\ and\ \citenamefont
  {Mlynek}}]{Felbinger}%
  \BibitemOpen
  \bibfield  {author} {\bibinfo {author} {\bibfnamefont {T.}~\bibnamefont
  {Felbinger}}, \bibinfo {author} {\bibfnamefont {S.}~\bibnamefont {Schiller}},
  \ and\ \bibinfo {author} {\bibfnamefont {J.}~\bibnamefont {Mlynek}},\ }\href
  {\doibase 10.1103/PhysRevLett.80.492} {\bibfield  {journal} {\bibinfo
  {journal} {Phys. Rev. Lett.}\ }\textbf {\bibinfo {volume} {80}},\ \bibinfo
  {pages} {492} (\bibinfo {year} {1998})}\BibitemShut {NoStop}%
\bibitem [{\citenamefont {Lindblad}(1976)}]{lindblad1976}%
  \BibitemOpen
  \bibfield  {author} {\bibinfo {author} {\bibfnamefont {G.}~\bibnamefont
  {Lindblad}},\ }\href {https://projecteuclid.org:443/euclid.cmp/1103899849}
  {\bibfield  {journal} {\bibinfo  {journal} {Comm. Math. Phys.}\ }\textbf
  {\bibinfo {volume} {48}},\ \bibinfo {pages} {119} (\bibinfo {year}
  {1976})}\BibitemShut {NoStop}%
\bibitem [{\citenamefont {Daley}(2014)}]{DaleyReview}%
  \BibitemOpen
  \bibfield  {author} {\bibinfo {author} {\bibfnamefont {A.~J.}\ \bibnamefont
  {Daley}},\ }\href {\doibase 10.1080/00018732.2014.933502} {\bibfield
  {journal} {\bibinfo  {journal} {Adv. Phys.}\ }\textbf {\bibinfo {volume}
  {63}},\ \bibinfo {pages} {77} (\bibinfo {year} {2014})}\BibitemShut {NoStop}%
\bibitem [{\citenamefont {Gardiner}\ and\ \citenamefont
  {Zoller}(2014)}]{QuantumOptics}%
  \BibitemOpen
  \bibfield  {author} {\bibinfo {author} {\bibfnamefont {C.~W.}\ \bibnamefont
  {Gardiner}}\ and\ \bibinfo {author} {\bibfnamefont {P.}~\bibnamefont
  {Zoller}},\ }\href@noop {} {\emph {\bibinfo {title} {Foundations of Quantum
  Optics}}},\ \bibinfo {edition} {2nd}\ ed.\ (\bibinfo  {publisher} {Imperial
  College Press},\ \bibinfo {year} {2014})\BibitemShut {NoStop}%
\bibitem [{\citenamefont {Plimak}\ \emph {et~al.}(2003)\citenamefont {Plimak},
  \citenamefont {Fleischhauer}, \citenamefont {Olsen},\ and\ \citenamefont
  {Collett}}]{StochasticDifferenceEquations}%
  \BibitemOpen
  \bibfield  {author} {\bibinfo {author} {\bibfnamefont {L.~I.}\ \bibnamefont
  {Plimak}}, \bibinfo {author} {\bibfnamefont {M.}~\bibnamefont
  {Fleischhauer}}, \bibinfo {author} {\bibfnamefont {M.~K.}\ \bibnamefont
  {Olsen}}, \ and\ \bibinfo {author} {\bibfnamefont {M.~J.}\ \bibnamefont
  {Collett}},\ }\href {\doibase 10.1103/PhysRevA.67.013812} {\bibfield
  {journal} {\bibinfo  {journal} {Phys. Rev. A}\ }\textbf {\bibinfo {volume}
  {67}},\ \bibinfo {pages} {013812} (\bibinfo {year} {2003})}\BibitemShut
  {NoStop}%
\bibitem [{\citenamefont {Olsen}\ \emph {et~al.}(2002)\citenamefont {Olsen},
  \citenamefont {Plimak},\ and\ \citenamefont
  {Fleischhauer}}]{ThreePhotonProcesses}%
  \BibitemOpen
  \bibfield  {author} {\bibinfo {author} {\bibfnamefont {M.~K.}\ \bibnamefont
  {Olsen}}, \bibinfo {author} {\bibfnamefont {L.~I.}\ \bibnamefont {Plimak}}, \
  and\ \bibinfo {author} {\bibfnamefont {M.}~\bibnamefont {Fleischhauer}},\
  }\href {\doibase 10.1103/PhysRevA.65.053806} {\bibfield  {journal} {\bibinfo
  {journal} {Phys. Rev. A}\ }\textbf {\bibinfo {volume} {65}},\ \bibinfo
  {pages} {053806} (\bibinfo {year} {2002})}\BibitemShut {NoStop}%
\bibitem [{\citenamefont {Drummond}\ and\ \citenamefont
  {Hillery}(2014)}]{drummond_quantum_2014}%
  \BibitemOpen
  \bibfield  {author} {\bibinfo {author} {\bibfnamefont {P.~D.}\ \bibnamefont
  {Drummond}}\ and\ \bibinfo {author} {\bibfnamefont {M.}~\bibnamefont
  {Hillery}},\ }\href {\doibase 10.1017/CBO9780511783616} {\emph {\bibinfo
  {title} {The {Quantum} {Theory} of {Nonlinear} {Optics}}}}\ (\bibinfo
  {publisher} {Cambridge University Press},\ \bibinfo {year}
  {2014})\BibitemShut {NoStop}%
\bibitem [{\citenamefont {Surya}\ \emph {et~al.}(2018)\citenamefont {Surya},
  \citenamefont {Guo}, \citenamefont {Zou},\ and\ \citenamefont
  {Tang}}]{surya_efficient_2018}%
  \BibitemOpen
  \bibfield  {author} {\bibinfo {author} {\bibfnamefont {J.~B.}\ \bibnamefont
  {Surya}}, \bibinfo {author} {\bibfnamefont {X.}~\bibnamefont {Guo}}, \bibinfo
  {author} {\bibfnamefont {C.-L.}\ \bibnamefont {Zou}}, \ and\ \bibinfo
  {author} {\bibfnamefont {H.~X.}\ \bibnamefont {Tang}},\ }\href {\doibase
  10.1364/OPTICA.5.000103} {\bibfield  {journal} {\bibinfo  {journal} {Optica}\
  }\textbf {\bibinfo {volume} {5}},\ \bibinfo {pages} {103} (\bibinfo {year}
  {2018})}\BibitemShut {NoStop}%
\bibitem [{\citenamefont {Surya}\ and\ \citenamefont
  {Tang}(2019)}]{surya_private_2019}%
  \BibitemOpen
  \bibfield  {author} {\bibinfo {author} {\bibfnamefont {J.~B.}\ \bibnamefont
  {Surya}}\ and\ \bibinfo {author} {\bibfnamefont {H.~X.}\ \bibnamefont
  {Tang}},\ }\href@noop {} {} (\bibinfo {year} {{private} communication)
  (2019})\BibitemShut {NoStop}%
\bibitem [{\citenamefont {Ning}\ \emph {et~al.}(2013)\citenamefont {Ning},
  \citenamefont {Hyvärinen}, \citenamefont {Pietarinen}, \citenamefont
  {Kaplas}, \citenamefont {Kauranen},\ and\ \citenamefont
  {Genty}}]{ning_third-harmonic_2013}%
  \BibitemOpen
  \bibfield  {author} {\bibinfo {author} {\bibfnamefont {T.}~\bibnamefont
  {Ning}}, \bibinfo {author} {\bibfnamefont {O.}~\bibnamefont {Hyvärinen}},
  \bibinfo {author} {\bibfnamefont {H.}~\bibnamefont {Pietarinen}}, \bibinfo
  {author} {\bibfnamefont {T.}~\bibnamefont {Kaplas}}, \bibinfo {author}
  {\bibfnamefont {M.}~\bibnamefont {Kauranen}}, \ and\ \bibinfo {author}
  {\bibfnamefont {G.}~\bibnamefont {Genty}},\ }\href {\doibase
  10.1364/OE.21.002012} {\bibfield  {journal} {\bibinfo  {journal} {Opt.
  Express}\ }\textbf {\bibinfo {volume} {21}},\ \bibinfo {pages} {2012}
  (\bibinfo {year} {2013})}\BibitemShut {NoStop}%
\bibitem [{\citenamefont {Mateen}\ \emph {et~al.}(2018)\citenamefont {Mateen},
  \citenamefont {Boales}, \citenamefont {Erramilli},\ and\ \citenamefont
  {Mohanty}}]{mateen_micromechanical_2018}%
  \BibitemOpen
  \bibfield  {author} {\bibinfo {author} {\bibfnamefont {F.}~\bibnamefont
  {Mateen}}, \bibinfo {author} {\bibfnamefont {J.}~\bibnamefont {Boales}},
  \bibinfo {author} {\bibfnamefont {S.}~\bibnamefont {Erramilli}}, \ and\
  \bibinfo {author} {\bibfnamefont {P.}~\bibnamefont {Mohanty}},\ }\href
  {\doibase 10.1038/s41378-018-0013-6} {\bibfield  {journal} {\bibinfo
  {journal} {Microsys. Nanoeng.}\ }\textbf {\bibinfo {volume} {4}},\ \bibinfo
  {pages} {14} (\bibinfo {year} {2018})}\BibitemShut {NoStop}%
\bibitem [{\citenamefont {Kischkat}\ \emph {et~al.}(2012)\citenamefont
  {Kischkat}, \citenamefont {Peters}, \citenamefont {Gruska}, \citenamefont
  {Semtsiv}, \citenamefont {Chashnikova}, \citenamefont {Klinkmüller},
  \citenamefont {Fedosenko}, \citenamefont {Machulik}, \citenamefont
  {Aleksandrova}, \citenamefont {Monastyrskyi}, \citenamefont {Flores},\ and\
  \citenamefont {Masselink}}]{kischkat_mid-infrared_2012}%
  \BibitemOpen
  \bibfield  {author} {\bibinfo {author} {\bibfnamefont {J.}~\bibnamefont
  {Kischkat}}, \bibinfo {author} {\bibfnamefont {S.}~\bibnamefont {Peters}},
  \bibinfo {author} {\bibfnamefont {B.}~\bibnamefont {Gruska}}, \bibinfo
  {author} {\bibfnamefont {M.}~\bibnamefont {Semtsiv}}, \bibinfo {author}
  {\bibfnamefont {M.}~\bibnamefont {Chashnikova}}, \bibinfo {author}
  {\bibfnamefont {M.}~\bibnamefont {Klinkmüller}}, \bibinfo {author}
  {\bibfnamefont {O.}~\bibnamefont {Fedosenko}}, \bibinfo {author}
  {\bibfnamefont {S.}~\bibnamefont {Machulik}}, \bibinfo {author}
  {\bibfnamefont {A.}~\bibnamefont {Aleksandrova}}, \bibinfo {author}
  {\bibfnamefont {G.}~\bibnamefont {Monastyrskyi}}, \bibinfo {author}
  {\bibfnamefont {Y.}~\bibnamefont {Flores}}, \ and\ \bibinfo {author}
  {\bibfnamefont {W.~T.}\ \bibnamefont {Masselink}},\ }\href {\doibase
  10.1364/AO.51.006789} {\bibfield  {journal} {\bibinfo  {journal} {Applied
  Optics}\ }\textbf {\bibinfo {volume} {51}},\ \bibinfo {pages} {6789}
  (\bibinfo {year} {2012})}\BibitemShut {NoStop}%
\bibitem [{\citenamefont {Rackauckas}\ and\ \citenamefont
  {Nie}(2016)}]{StocDiffEq}%
  \BibitemOpen
  \bibfield  {author} {\bibinfo {author} {\bibfnamefont {C.}~\bibnamefont
  {Rackauckas}}\ and\ \bibinfo {author} {\bibfnamefont {Q.}~\bibnamefont
  {Nie}},\ }\href {\doibase 10.3934/dcdsb.2017133} {\bibfield  {journal}
  {\bibinfo  {journal} {Discrete Continuous Dyn. Syst. Ser. B}\ }\textbf
  {\bibinfo {volume} {22}},\ \bibinfo {pages} {2731} (\bibinfo {year}
  {2016})}\BibitemShut {NoStop}%
\bibitem [{\citenamefont {Bezanson}\ \emph {et~al.}(2017)\citenamefont
  {Bezanson}, \citenamefont {Edelman}, \citenamefont {Karpinski},\ and\
  \citenamefont {Shah}}]{Bezanson:2017gd}%
  \BibitemOpen
  \bibfield  {author} {\bibinfo {author} {\bibfnamefont {J.}~\bibnamefont
  {Bezanson}}, \bibinfo {author} {\bibfnamefont {A.}~\bibnamefont {Edelman}},
  \bibinfo {author} {\bibfnamefont {S.}~\bibnamefont {Karpinski}}, \ and\
  \bibinfo {author} {\bibfnamefont {V.~B.}\ \bibnamefont {Shah}},\ }\href
  {\doibase 10.1137/141000671} {\bibfield  {journal} {\bibinfo  {journal} {SIAM
  Review}\ }\textbf {\bibinfo {volume} {59}},\ \bibinfo {pages} {65} (\bibinfo
  {year} {2017})}\BibitemShut {NoStop}%
\bibitem [{\citenamefont {Rackauckas}\ and\ \citenamefont
  {Nie}(2017)}]{DiffEq}%
  \BibitemOpen
  \bibfield  {author} {\bibinfo {author} {\bibfnamefont {C.}~\bibnamefont
  {Rackauckas}}\ and\ \bibinfo {author} {\bibfnamefont {Q.}~\bibnamefont
  {Nie}},\ }\href {\doibase 10.5334/jors.151} {\bibfield  {journal} {\bibinfo
  {journal} {J. Open Source Softw.}\ }\textbf {\bibinfo {volume} {5}},\
  \bibinfo {pages} {15} (\bibinfo {year} {2017})}\BibitemShut {NoStop}%
\bibitem [{\citenamefont {Olsen}\ and\ \citenamefont
  {Bradley}(2009)}]{sampling}%
  \BibitemOpen
  \bibfield  {author} {\bibinfo {author} {\bibfnamefont {M.~K.}\ \bibnamefont
  {Olsen}}\ and\ \bibinfo {author} {\bibfnamefont {A.~S.}\ \bibnamefont
  {Bradley}},\ }\href {\doibase 10.1016/j.optcom.2009.06.033} {\bibfield
  {journal} {\bibinfo  {journal} {Opt. Commun.}\ }\textbf {\bibinfo {volume}
  {282}},\ \bibinfo {pages} {3924} (\bibinfo {year} {2009})}\BibitemShut
  {NoStop}%
\bibitem [{\citenamefont {Olsen}(2004)}]{SHG}%
  \BibitemOpen
  \bibfield  {author} {\bibinfo {author} {\bibfnamefont {M.~K.}\ \bibnamefont
  {Olsen}},\ }\href {\doibase 10.1103/PhysRevA.70.035801} {\bibfield  {journal}
  {\bibinfo  {journal} {Phys. Rev. A}\ }\textbf {\bibinfo {volume} {70}},\
  \bibinfo {pages} {035801} (\bibinfo {year} {2004})}\BibitemShut {NoStop}%
\bibitem [{\citenamefont {Gardiner}(2009)}]{StochasticMethods}%
  \BibitemOpen
  \bibfield  {author} {\bibinfo {author} {\bibfnamefont {C.~W.}\ \bibnamefont
  {Gardiner}},\ }\href@noop {} {\emph {\bibinfo {title} {Handbook of Stochasitc
  Methods}}},\ \bibinfo {edition} {4th}\ ed.\ (\bibinfo  {publisher}
  {Springer},\ \bibinfo {year} {2009})\BibitemShut {NoStop}%
\bibitem [{\citenamefont {Chaturvedi}\ and\ \citenamefont
  {Gardiner}(1977)}]{Gardiner1977}%
  \BibitemOpen
  \bibfield  {author} {\bibinfo {author} {\bibfnamefont {S.}~\bibnamefont
  {Chaturvedi}}\ and\ \bibinfo {author} {\bibfnamefont {C.~W.}\ \bibnamefont
  {Gardiner}},\ }\href {\doibase 10.1007/BF01014349} {\bibfield  {journal}
  {\bibinfo  {journal} {J. Stat. Phys.}\ }\textbf {\bibinfo {volume} {17}},\
  \bibinfo {pages} {429} (\bibinfo {year} {1977})}\BibitemShut {NoStop}%
\bibitem [{\citenamefont {Walls}\ and\ \citenamefont
  {Milburn}(1994)}]{WallsMilburn}%
  \BibitemOpen
  \bibfield  {author} {\bibinfo {author} {\bibfnamefont {D.~F.}\ \bibnamefont
  {Walls}}\ and\ \bibinfo {author} {\bibfnamefont {G.~J.}\ \bibnamefont
  {Milburn}},\ }\href@noop {} {\emph {\bibinfo {title} {Quantum Optics}}}\
  (\bibinfo  {publisher} {Springer},\ \bibinfo {year} {1994})\BibitemShut
  {NoStop}%
\bibitem [{\citenamefont {Gardiner}\ and\ \citenamefont
  {Collett}(1985)}]{InputOutput}%
  \BibitemOpen
  \bibfield  {author} {\bibinfo {author} {\bibfnamefont {C.~W.}\ \bibnamefont
  {Gardiner}}\ and\ \bibinfo {author} {\bibfnamefont {M.~J.}\ \bibnamefont
  {Collett}},\ }\href {\doibase 10.1103/PhysRevA.31.3761} {\bibfield  {journal}
  {\bibinfo  {journal} {Phys. Rev. A}\ }\textbf {\bibinfo {volume} {31}},\
  \bibinfo {pages} {3761} (\bibinfo {year} {1985})}\BibitemShut {NoStop}%
\bibitem [{\citenamefont {Caves}(1981)}]{Caves}%
  \BibitemOpen
  \bibfield  {author} {\bibinfo {author} {\bibfnamefont {C.~M.}\ \bibnamefont
  {Caves}},\ }\href {\doibase 10.1103/PhysRevD.23.1693} {\bibfield  {journal}
  {\bibinfo  {journal} {Phys. Rev. D}\ }\textbf {\bibinfo {volume} {23}},\
  \bibinfo {pages} {1693} (\bibinfo {year} {1981})}\BibitemShut {NoStop}%
\bibitem [{\citenamefont {Walls}(1983)}]{walls_squeezed_1983}%
  \BibitemOpen
  \bibfield  {author} {\bibinfo {author} {\bibfnamefont {D.~F.}\ \bibnamefont
  {Walls}},\ }\href {\doibase https://doi.org/10.1038/306141a0} {\bibfield
  {journal} {\bibinfo  {journal} {Nature}\ }\textbf {\bibinfo {volume} {306}},\
  \bibinfo {pages} {141} (\bibinfo {year} {1983})}\BibitemShut {NoStop}%
\bibitem [{\citenamefont {Duan}\ \emph {et~al.}(2000)\citenamefont {Duan},
  \citenamefont {Giedke}, \citenamefont {Cirac},\ and\ \citenamefont
  {Zoller}}]{Duan}%
  \BibitemOpen
  \bibfield  {author} {\bibinfo {author} {\bibfnamefont {L.-M.}\ \bibnamefont
  {Duan}}, \bibinfo {author} {\bibfnamefont {G.}~\bibnamefont {Giedke}},
  \bibinfo {author} {\bibfnamefont {J.~I.}\ \bibnamefont {Cirac}}, \ and\
  \bibinfo {author} {\bibfnamefont {P.}~\bibnamefont {Zoller}},\ }\href
  {\doibase 10.1103/PhysRevLett.84.2722} {\bibfield  {journal} {\bibinfo
  {journal} {Phys. Rev. Lett.}\ }\textbf {\bibinfo {volume} {84}},\ \bibinfo
  {pages} {2722} (\bibinfo {year} {2000})}\BibitemShut {NoStop}%
\bibitem [{\citenamefont {Simon}(2000)}]{Simon}%
  \BibitemOpen
  \bibfield  {author} {\bibinfo {author} {\bibfnamefont {R.}~\bibnamefont
  {Simon}},\ }\href {\doibase 10.1103/PhysRevLett.84.2726} {\bibfield
  {journal} {\bibinfo  {journal} {Phys. Rev. Lett.}\ }\textbf {\bibinfo
  {volume} {84}},\ \bibinfo {pages} {2726} (\bibinfo {year}
  {2000})}\BibitemShut {NoStop}%
\bibitem [{\citenamefont {Bradley}\ \emph {et~al.}(2005)\citenamefont
  {Bradley}, \citenamefont {Olsen}, \citenamefont {Pfister},\ and\
  \citenamefont {Pooser}}]{BrightTripartite}%
  \BibitemOpen
  \bibfield  {author} {\bibinfo {author} {\bibfnamefont {A.~S.}\ \bibnamefont
  {Bradley}}, \bibinfo {author} {\bibfnamefont {M.~K.}\ \bibnamefont {Olsen}},
  \bibinfo {author} {\bibfnamefont {O.}~\bibnamefont {Pfister}}, \ and\
  \bibinfo {author} {\bibfnamefont {R.~C.}\ \bibnamefont {Pooser}},\ }\href
  {\doibase 10.1103/PhysRevA.72.053805} {\bibfield  {journal} {\bibinfo
  {journal} {Phys. Rev. A}\ }\textbf {\bibinfo {volume} {72}},\ \bibinfo
  {pages} {053805} (\bibinfo {year} {2005})}\BibitemShut {NoStop}%
\bibitem [{\citenamefont {Dalibard}\ \emph {et~al.}(1992)\citenamefont
  {Dalibard}, \citenamefont {Castin},\ and\ \citenamefont
  {M\o{}lmer}}]{MCWF-Dalibard}%
  \BibitemOpen
  \bibfield  {author} {\bibinfo {author} {\bibfnamefont {J.}~\bibnamefont
  {Dalibard}}, \bibinfo {author} {\bibfnamefont {Y.}~\bibnamefont {Castin}}, \
  and\ \bibinfo {author} {\bibfnamefont {K.}~\bibnamefont {M\o{}lmer}},\ }\href
  {\doibase 10.1103/PhysRevLett.68.580} {\bibfield  {journal} {\bibinfo
  {journal} {Phys. Rev. Lett.}\ }\textbf {\bibinfo {volume} {68}},\ \bibinfo
  {pages} {580} (\bibinfo {year} {1992})}\BibitemShut {NoStop}%
\bibitem [{\citenamefont {Dum}\ \emph {et~al.}(1992)\citenamefont {Dum},
  \citenamefont {Zoller},\ and\ \citenamefont {Ritsch}}]{MCWF-Dum}%
  \BibitemOpen
  \bibfield  {author} {\bibinfo {author} {\bibfnamefont {R.}~\bibnamefont
  {Dum}}, \bibinfo {author} {\bibfnamefont {P.}~\bibnamefont {Zoller}}, \ and\
  \bibinfo {author} {\bibfnamefont {H.}~\bibnamefont {Ritsch}},\ }\href
  {\doibase 10.1103/PhysRevA.45.4879} {\bibfield  {journal} {\bibinfo
  {journal} {Phys. Rev. A}\ }\textbf {\bibinfo {volume} {45}},\ \bibinfo
  {pages} {4879} (\bibinfo {year} {1992})}\BibitemShut {NoStop}%
\bibitem [{\citenamefont {Krämer}\ \emph {et~al.}(2018)\citenamefont
  {Krämer}, \citenamefont {Plankensteiner}, \citenamefont {Ostermann},\ and\
  \citenamefont {Ritsch}}]{QO}%
  \BibitemOpen
  \bibfield  {author} {\bibinfo {author} {\bibfnamefont {S.}~\bibnamefont
  {Krämer}}, \bibinfo {author} {\bibfnamefont {D.}~\bibnamefont
  {Plankensteiner}}, \bibinfo {author} {\bibfnamefont {L.}~\bibnamefont
  {Ostermann}}, \ and\ \bibinfo {author} {\bibfnamefont {H.}~\bibnamefont
  {Ritsch}},\ }\href {\doibase https://doi.org/10.1016/j.cpc.2018.02.004}
  {\bibfield  {journal} {\bibinfo  {journal} {Comput. Phys. Commun.}\ }\textbf
  {\bibinfo {volume} {227}},\ \bibinfo {pages} {109 } (\bibinfo {year}
  {2018})}\BibitemShut {NoStop}%
\end{thebibliography}

%

\end{document}